\documentclass[pra,twocolumn,showpacs,superscriptaddress,floatfix]{revtex4-1}
\usepackage{graphicx}
\usepackage{epstopdf} 
\usepackage{amsmath}
\usepackage{epsfig}
\usepackage{latexsym}
\usepackage{amsfonts}
\usepackage{graphics}
\usepackage{bm}
\usepackage{braket}
\usepackage{dsfont}

\usepackage{mathtools}

\usepackage{helvet}

\begin{document}

\date{\today}
\author{J. Mumford}
\affiliation{Department of Physics and Astronomy, McMaster University, 1280 Main St.\ W., Hamilton, ON, L8S 4M1, Canada}
\affiliation{School of Arts and Sciences, Red Deer College, 100 College Boulevard, Red Deer, AB, T4N 5H5, Canada}
\author{W. Kirkby}
\affiliation{Department of Physics and Astronomy, McMaster University, 1280 Main St.\ W., Hamilton, ON, L8S 4M1, Canada}
\author{D. H. J. O'Dell}
\affiliation{Department of Physics and Astronomy, McMaster University, 1280 Main St.\ W., Hamilton, ON, L8S 4M1, Canada}

\title{Measuring out-of-time-ordered correlation functions with a single impurity qubit in a bosonic Josephson junction}

\begin{abstract}
We calculate the out-of-time-ordered correlation function (OTOC) of a single impurity qubit coupled to fully a connected many-particle system such as a bosonic Josephson junction or spins with long-range interactions. In these systems the qubit OTOC can be used to detect both ground state and excited state quantum phase transitions (QPTs), making it a robust order parameter that is considerably more sensitive than the standard one-body correlation function.  Finite size scaling exponents for an $N$ body system can also be  accurately extracted from  the long-time OTOC dynamics, however, for short times there is a discrepancy due to the fact that the qubit has not had enough time to couple to the larger system.   Our results show that the OTOC of even the smallest probe is enough to diagnose a QPT in fully connected models but, like a continuous measurement, can still cause a backaction effect which leads to weakly chaotic dynamics and gradual information scrambling. 
\end{abstract}

\maketitle

\section{Introduction}


Experimental progress in recent decades has led to the realization of highly isolated quantum systems which can evolve unitarily
over time scales which are long compared to a single natural cycle of the system \cite{bloch08,blatt12,georgescu14}.  This has allowed the observation of many-particle quantum dynamical phenomena such as far from equilibrium universal dynamics \cite{prufer18}, discrete time crystals \cite{zhang17,choi17}, and ultracold atom analogues of Sakharov oscillations \cite{Hung19} and black holes \cite{steinhauer14}.  It has recently become clear that a particularly useful tool in the study of the dynamics of many-particle quantum systems is a four-point correlation function known as the out-of-time-ordered correlation function (OTOC).  Theoretical work has shown OTOCs to be excellent measures of information scrambling \cite{yao18,swingle16,bohrdt17,swan18} in chaotic systems \cite{roberts15,zhu16,kukuljan17,rozenbaum17,cotler18,rozenbaum18,kurchan18,chen18,mata18,carlos18,jalabert18,hamazaki18,herrera18} where the scrambling rate can be related to the Lyapunov exponent.  It has also been pointed out that OTOCs are capable of identifying many-body localization \cite{he17,chen17,huang17,dag18} and entanglement growth \cite{fan17,hosur16}. 

Yet another application of OTOCs is through their long-time averages which act as sensitive order parameters that are able to diagnose quantum phase transitions (QPTs). It is this latter feature which is the main focus of this paper; prior work has shown how information about QPTs can be extracted from OTOCs  calculated for the transverse field Ising model (TFIM) and the Lipkin-Meshkov-Glick model (LMG) \cite{heyl18} and also the Dicke model \cite{sun19}.  Our main innovation here is to consider the OTOC for a single subunit of the total system, i.e.\ a single ``impurity'' qubit, which is picked out and addressed. We find that the OTOC of even a single such qubit contains the necessary information to identify a QPT in the larger system. 

The general form of an OTOC is 
\begin{equation}
F(t)=\langle \hat{B}(t)^\dagger \hat{A}(0)^\dagger \hat{B}(t) \hat{A}(0) \rangle
\label{eq:otoc} 
\end{equation}
where $\hat{B}(t) = e^{i \hat{H} t} \hat{B}(0) e^{-i \hat{H}t}$ and the expectation value $\langle ... \rangle$ is typically performed using a pure state or a thermal average. We shall only be concerned with the former case in this paper. The operators are chosen such that they initially commute, $[ \hat{B}(0), \hat{A}(0)] = 0$.  Some intuition can be gained about Eq.\ \eqref{eq:otoc} by considering it to be the overlap between two states, $F(t) = \langle \psi_2 (t)\vert \psi_1(t) \rangle$, where $\vert \psi_1 (t) \rangle = \hat{B}(t) \hat{A}(0) \vert \psi \rangle$ and $\vert \psi_2 (t) \rangle = \hat{A}(0) \hat{B}(t) \vert \psi \rangle$.  When this overlap decays exponentially the situation becomes reminiscent of classical chaos where there is an exponential separation of two classical trajectories in phase space due to a small initial difference. Recent studies on OTOCs have been able to identify Lyapunov exponents and address the connection between information scrambling in quantum systems and black hole physics \cite{maldacena16,shen17}. In the OTOC the perturbation is the non-commutativity of $\hat{A}(0)$ and $\hat{B}(t)$ for $t>0$.  This can be shown explicitly when $\hat{A}$ and $\hat{B}$ are both Hermitian and unitary; the real part of the OTOC function then takes the form 
\begin{equation}
\mathrm{Re}[F(t)] = 1 - C(t)/2
\end{equation} 
where $C(t) = - \braket{[\hat{A}(0), \hat{B}(t) ]^2 }$ is the out-of-time-ordered commutator.

If OTOCs are to be a truly useful concept they should be experimentally accessible, and, indeed, OTOCs have been successfully measured in various systems \cite{wei18,garttner17,li17,landsman19,meier19}. However, challenges remain because of the many processes implied by Eq.\ \eqref{eq:otoc}.  The preparation of state $\vert \psi_{1} \rangle$ requires the following four steps: (1) application of operator $\hat{A}$, (2) forward time propagation, (3) application of operator $\hat{B}$, (4) backward time propagation.  The preparation of state $\vert \psi_2 \rangle$ requires the same number of steps.  To simplify matters, proposals have been put forward for obtaining OTOCs through measurements on simple ancillary systems such as harmonic oscillators \cite{chaudhuri19} or qubits \cite{swingle16}. In the latter case this results in the entangled state of the form $\left ( \vert \psi_1(t) \rangle \vert + \rangle_Q + \vert \psi_2 (t) \rangle \vert - \rangle_Q \right )/2$, so that the OTOC can be obtained through measurements of the $x$-component of the qubit's spin, $\mathrm{Re}\left [ F(t) \right ] = \langle \hat{\sigma}_x \rangle$.  Although, this method has the benefit of producing no backaction from the qubit onto the system, it requires a five step gate sequence to create the above entangled state.  We show that when the qubit is not ancillary, but instead coupled to the system via contact interactions the OTOCs can be obtained from measurements of the qubit without a gate sequence.  Our method results in backaction from the qubit, however, we find that it does not diminish the qubit's ability to act as a probe of the QPT in the many-particle system.  Furthermore, the backaction from the qubit is interesting in itself because it has been shown that it is a source of classical chaos in our system \cite{mulansky11}.  As the exponential decay of Eq.\ \eqref{eq:otoc} resembles the separation of trajectories in a classically chaotic system, we take this opportunity to use the OTOCs to measure the amount of chaos the qubit introduces to the system.  We show that when a proper state is chosen Eq.\ \eqref{eq:otoc} simplifies considerably opening up the possibility of additional ways to experimentally observe how chaos is introduced into many-body systems.  


The quantum many-particle systems we have in mind in this paper are trapped cold atom or ion setups. Trapped ions are intrinsically amenable to the realization of single particle probes because the typical separation between ions is large enough that they can easily be resolved optically. This allows for individual ion addressing and read out such that the spin state of an ion can be determined with an error of less than $10^{-3}$  \cite{Harty14} (quantum state tomography, which goes beyond a simple projective measurement, is also a well developed technique in ions \cite{Lanyon17}.) Neutral atomic gases tend to be denser and hence individual addressing is more challenging, but nevertheless single atom microscopes have been developed which, for example, have been used to probe the superfluid-to-Mott insulator QPT in an optical lattice \cite{bakr10, sherson10}.  There have also been experiments involving single ions immersed  in $^{87}\mathrm{Rb}$ Bose-Einstein condensates (BECs) demonstrating that the ion can both be controlled independently of the BEC \cite{zipkes10} and act as a probe of the BEC's density profile \cite{schmid10}. The ultracold regime where the ion-atom collisions are in the $s$-wave channel is now being approached \cite{Pinkas19}.  Single neutral impurity atoms of $^{133}\mathrm{Cs}$ have also been successfully immersed in $^{87}\mathrm{Rb}$ BECs to track three-body collisions \cite{spethmann12}, spin exchange between the BEC and impurity  \cite{schmid18}, and  also  monitor interaction times \cite{hohmann17}.  Furthermore, proposals involving qubit probes have been put forward showing that the dynamics of a qubit coupled to a single site of a 2D lattice can encode information about the local excitation spectrum \cite{usui18} and the dephasing of a qubit in an ultracold atomic gas can encode information about the density and spatial variance of the gas \cite{elliott16}.

\section{Model}
\label{sec:model}

Our model consists of $N$ identical two-level interacting bosons coupled to a single impurity qubit. We can write the total Hamiltonian as
\begin{equation}
	\hat{H} = \hat{H}_B + \hat{H}_Q + \hat{H}_I
	\label{eq:ham}
\end{equation}
where $\hat{H}_B$, $\hat{H}_Q$ and $\hat{H}_I$ are the $N$ boson, qubit, and interaction Hamiltonians, respectively, and are given by
\begin{eqnarray}
	\hat{H}_B &=& U \hat{S}_z^2 - 2 J \hat{S}_x \nonumber \\
	\hat{H}_Q &=& - N J^a \hat{\sigma}_x \nonumber \\
	\hat{H}_I &=& W \hat{S}_z \hat{\sigma}_z \, .
\end{eqnarray}
The collective spin operators are defined as sums of individual Pauli operators for the bosons, $\hat{S}_\alpha = 1/2 \sum_i^N \hat{\sigma}_\alpha^i$, where $\alpha \in\{x, y, z\}$ (units of $\hbar=1$).  The qubit operators are also Pauli matrices, however, to distinguish them the superscript label has been removed. The parameters in the system are defined in the following way: $U$ is the boson-boson interaction energy, $J$ is the boson transition energy, $J^a$ is the qubit transition energy, and $W$ is the boson-qubit interaction energy.  The introduction of $N$ in the qubit term is known as the \textit{Kac prescription} \cite{kac63} and is applied in order for $\hat{H}$ to scale appropriately in the thermodynamic limit, $N \to \infty$.  

This collective spin model can be used to describe an ensemble of $N$ trapped ions each of which has two relevant internal states (the two spin states) and which have all-to-all interactions. Tunable long-range interactions between ions have been demonstrated in traps containing up to hundreds of ions where interactions are mediated by phonons and controlled by lasers  \cite{Islam11,Britton12,Islam13,Jurcevic14,Richerme14,Bohnet16}.  The inter-ion interaction potential in these systems has the form $1/ \vert \mathbf{r}-\mathbf{r}' \vert^{\epsilon}$ where $\epsilon$ can be as small as 0.02 \cite{Bohnet16}, thus making the interaction largely independent of position (the mapping of such systems to our Hamiltonian can be found in reference \cite{Das06}). Equally, the same collective spin model can be realized using scalar bosonic atoms trapped in a double well potential \cite{albeiz05,levy2007,Trenkwalder16} such that we realize the two-mode Bose-Hubbard model in which case the spin labels refer to the left or right well, or by spinor bosonic atoms trapped in the same mode of a single well \cite{Zibold10}.  In all three cases the impurity could correspond to one of the atoms/ions that is distinguishable in some way, e.g.\ by being a different species, or by being in a different set of internal states. The model is trivially extendable to the case of $M$ impurities providing they only significantly interact with the bosons and not each other, e.g.\ a low density of impurities. 

For the sake of developing intuition, let us consider a bosonic Josephson junction which consists of ultracold scalar bosons inhabiting a double well potential; the height of the barrier determines $J$. The impurity inhabits a double well which overlaps with that of the bosons although $J^{a}$ can in general be different. When the magnitude of $W$ exceeds a certain critical value, $W_{c}$, the system undergoes a QPT which spontaneously breaks the symmetry of the Hamiltonian. If $W$ is positive (signifying repulsive boson-impurity interactions) the bosons clump in one well and expel the impurity to the other \cite{Rinck2011}.  If $W$ is negative (signifying attractive boson-impurity interactions) the bosons and impurity choose the same well. As shown in references \cite{mumford14a} and \cite{mumford14b}, this QPT is a $\mathbb{Z}_2$ symmetry-breaking transition similar to the QPT in the celebrated Dicke model and in fact falls into the same universality class.  

We recall that the Dicke model  describes $N$ two-level atoms collectively coupled to a single mode of the electromagnetic field represented by a harmonic oscillator; for small values of the atom-field coupling the ground state of the whole system has the harmonic oscillator in its ground state (paramagnetic/normal state) but above a critical value  there is a transition to a superradiant (ferromagnetic/symmetry broken) state where the field spontaneously becomes excited \cite{Dicke54,Hepp73,Wang73,Garraway11,Dimer07,Nagy10,Bhaseen12}. Recent realizations of this transition include both an actual superradiant transition in an optical cavity filled with an atomic BEC pumped from the side by a laser \cite{Baumann10}, and simulations in trapped ion systems \cite{Safavi18}. In our case, the two levels of the impurity simulate the first two levels of the harmonic oscillator and, remarkably, this is enough to capture the critical properties of the Dicke model which are determined by the regime where the electromagnetic field is barely excited \cite{mumford14b}. Another similarity between the Dicke model and the boson-impurity model is that both display regular dynamics in the normal phase and chaotic dynamics in the symmetry broken phase \cite{mumford14a,lambert04}. We note in this context that there are regimes where the boson-impurity model can be mapped onto the double-pendulum Hamiltonian which is a prototypical chaotic system \cite{mulansky11}.

In fact, even without the impurity the bosons can break the $\mathbb{Z}_2$ symmetry of $\hat{H}_{B}$
 if the self-interaction $U$ is made attractive (negative) enough such that they bunch into a single well  \cite{buonsante12}. The dimensionless  parameter controlling the transition is the ratio of the self-interaction energy to the boson hopping energy 
\begin{equation}
\Lambda \equiv \frac{UN}{2J} \ ,
\end{equation}
and the quantum critical point is $\Lambda_c = -1$. To understand what the order parameter should be for this transition  it is useful to again turn to the double well model and use the Schwinger representation \cite{Sakurai93} to re-express the spin operators in terms of bosonic annihilation/creation operators for the left $\hat{b}_{l}/\hat{b}^{\dag}_{l}$ and right $\hat{b}_{r}/\hat{b}^{\dag}_{r}$ wells or modes 
\begin{eqnarray}
\hat{S}_{z} & = & (\hat{b}_{l}^{\dag}\hat{b}_{l}-\hat{b}_{r}^{\dag}\hat{b}_{r})/2 \\
\hat{S}_{x} & = & (\hat{b}_{r}^{\dag}\hat{b}_{l}+\hat{b}_{l}^{\dag}\hat{b}_{r})/2 \ .
\end{eqnarray}
Exactly analogous expressions can be written down for the impurity.
Clearly, $\hat{S}_{z}$ is the operator corresponding to half the difference in the number of bosons between the left and right modes. It has eigenvalues lying in the range $\pm N/2$ and its eigenvectors form a complete basis for describing the bosonic many particle wave function.  $\hat{S}_{x}$ is the hopping operator which also gives the coherence between the two modes \cite{Gati07}. It has the same eigenvalues as  $\hat{S}_{z}$ and its eigenvectors form an alternative complete basis. Consider first the probability distribution for the quantum mechanical ground state in the $\hat{S}_{z}$ basis: this is \textit{always} symmetric about zero so that if $\hat{S}_{z}$ is measured it will randomly give a positive or negative value with equal probability so that when averaged over many experimental runs we find $\langle \hat{S}_{z} \rangle=0$. The only difference below and above the transition is that the probability distribution below has a single peak centred at zero and above develops two peaks symmetrically distributed about zero so that as $U$ is made more and more negative it becomes a Schr\"{o}dinger cat state \cite{mulansky11}. Therefore the single particle correlator $\langle \hat{S_{z}} \rangle$ 
is insensitive to the transition and is not a good order parameter.

Now consider the situation in the $\hat{S}_{x}$ basis. Introducing the symmetric $\hat{b}_{s}=(\hat{b}_{l}+\hat{b}_{r})/\sqrt{2}$ and antisymmetric $\hat{b}_{a}=(\hat{b}_{l}-\hat{b}_{r})/\sqrt{2}$ combinations of the left and right mode operators, $\hat{S}_{x}$ can be alternatively written as
\begin{equation}
\hat{S}_{x}=(\hat{b}_{s}^{\dag}\hat{b}_{s}-\hat{b}_{a}^{\dag}\hat{b}_{a})/2 \ .
\end{equation}
Thus, $\hat{S}_{x}$ gives half the difference in number between the  symmetric and antisymmetric modes. However, unlike $\hat{S}_{z}$, the probability distribution for the ground state wave function in the  $\hat{S}_{x}$ basis is not symmetric about zero: if $U=0$ all the bosons occupy the lower energy symmetric mode and $\langle \hat{S}_{x} \rangle=-N/2$. Finite values of $U$ excite particles into the antisymmetric mode and as $\vert U \vert \rightarrow \infty$ the two modes become equally occupied so that $\langle \hat{S}_{x} \rangle=0$. In the thermodynamic limit $N \rightarrow \infty$ the QPT becomes sharp such that $\langle \hat{S}_{x} \rangle=0$ for $\Lambda > \Lambda_{c}$ (normal phase) and takes finite values for $\Lambda < \Lambda_{c}$ ($\mathbb{Z}_2$ symmetry broken phase) \cite{mumford14a,mumford14b,buonsante12}. Therefore $\langle \hat{S_{x}} \rangle$ 
is a good order parameter for this transition.

 Importantly, the introduction of the two qubit related terms does not break the symmetry of $\hat{H}_B$, but instead shifts the quantum critical point, giving \cite{mulansky11}
\begin{equation}
\Lambda_c = \frac{W^2}{4 J^a} - 1\;.
\label{eq:critlam}
\end{equation}
From this point forward, energies ($J^a$, $W$, etc.) will be given in units of the boson tunneling parameter, $J$. Apart from the Dicke model, this QPT is also in the same universality class as that of the closely related LMG model  \cite{lipkin65,dusuel04} and the infinite range transverse field Ising model \cite{Das06}, where in the latter case it corresponds to a transition between paramagnetic and ferromagnetic phases.  

%



\section{Using a qubit to sense a QPT}

\begin{figure}[t!]
        \includegraphics[height=3.2cm]{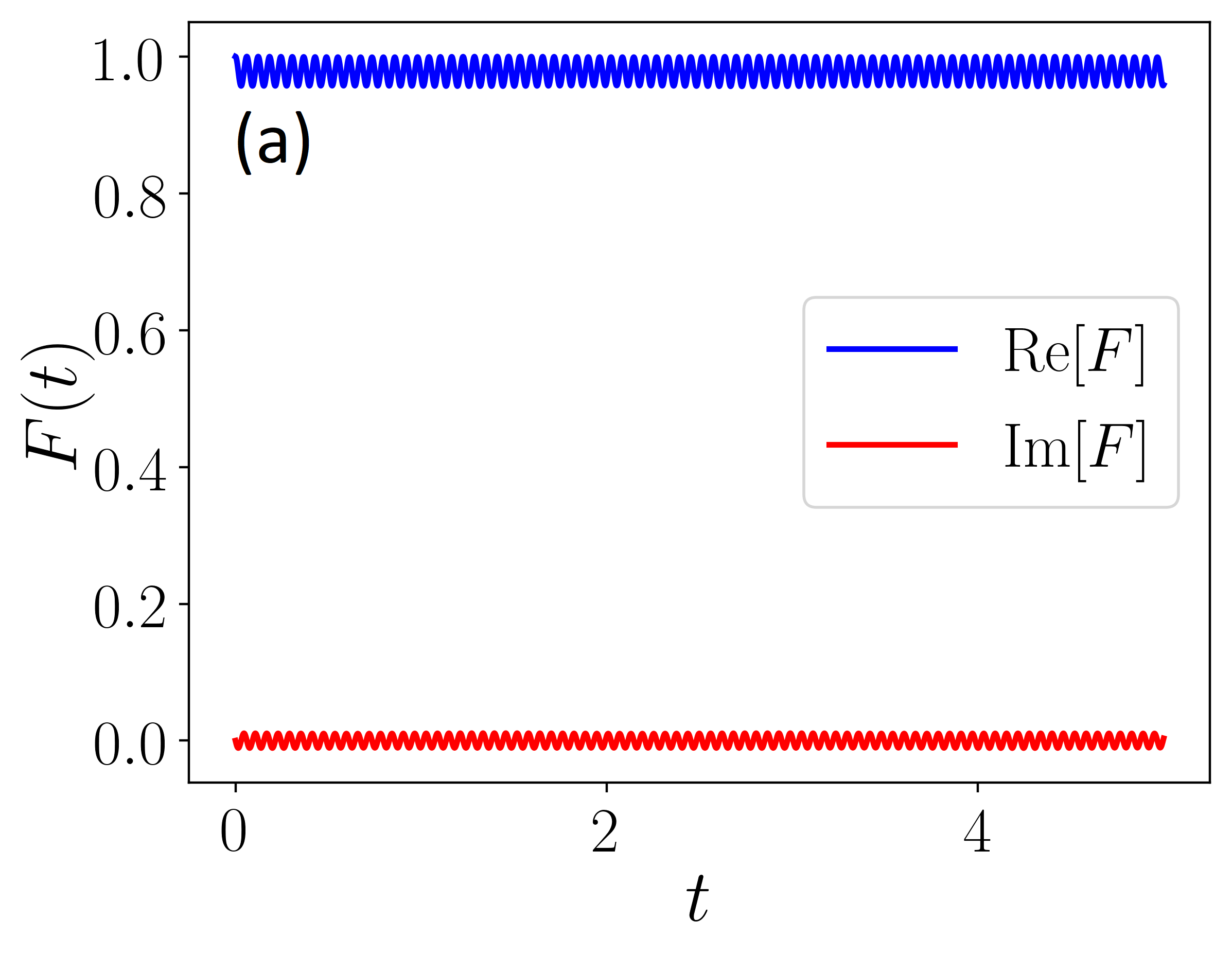} 
        \includegraphics[height=3.2cm]{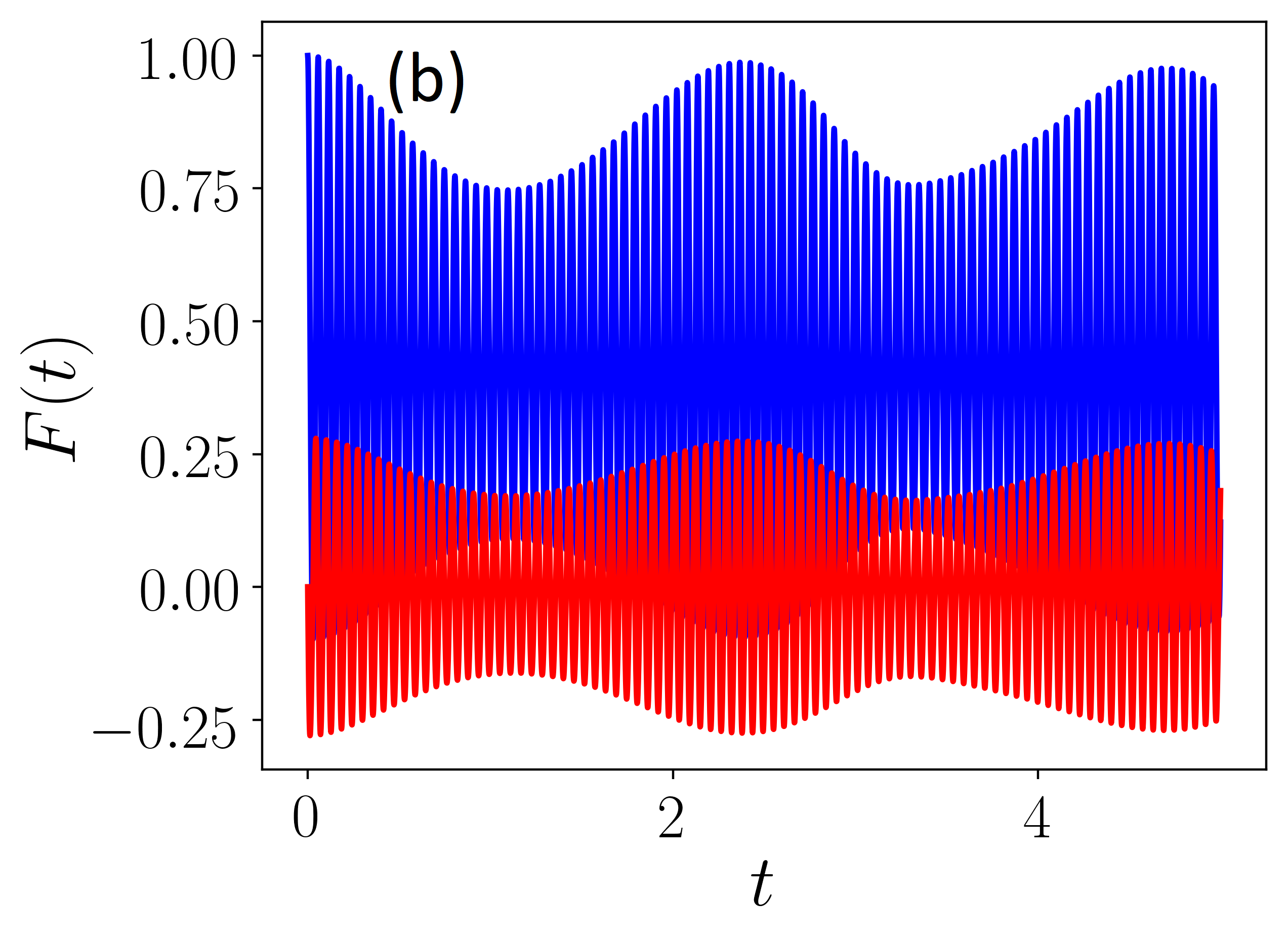} 
        \caption{The time dependence of the qubit OTOC on either side of the QPT: (a) the normal phase and (b) the symmetry broken phase.  Each image shows $\mathrm{Re}[F(t)]$ (blue) and $\mathrm{Im}[F(t)]$ (red).  The imaginary component of the OTOC function oscillates around zero in both phases, but the real part oscillates around unity in the normal phase and less than unity in the symmetry broken phase.  The parameters used are $J^a = W = 1.0$ (in units of $J$) and $N = 50$ giving $\Lambda_c = -3/4$.}
\label{fig:otoccol}
\end{figure}

\begin{figure*}
        \includegraphics[height=4.5cm,width=6cm]{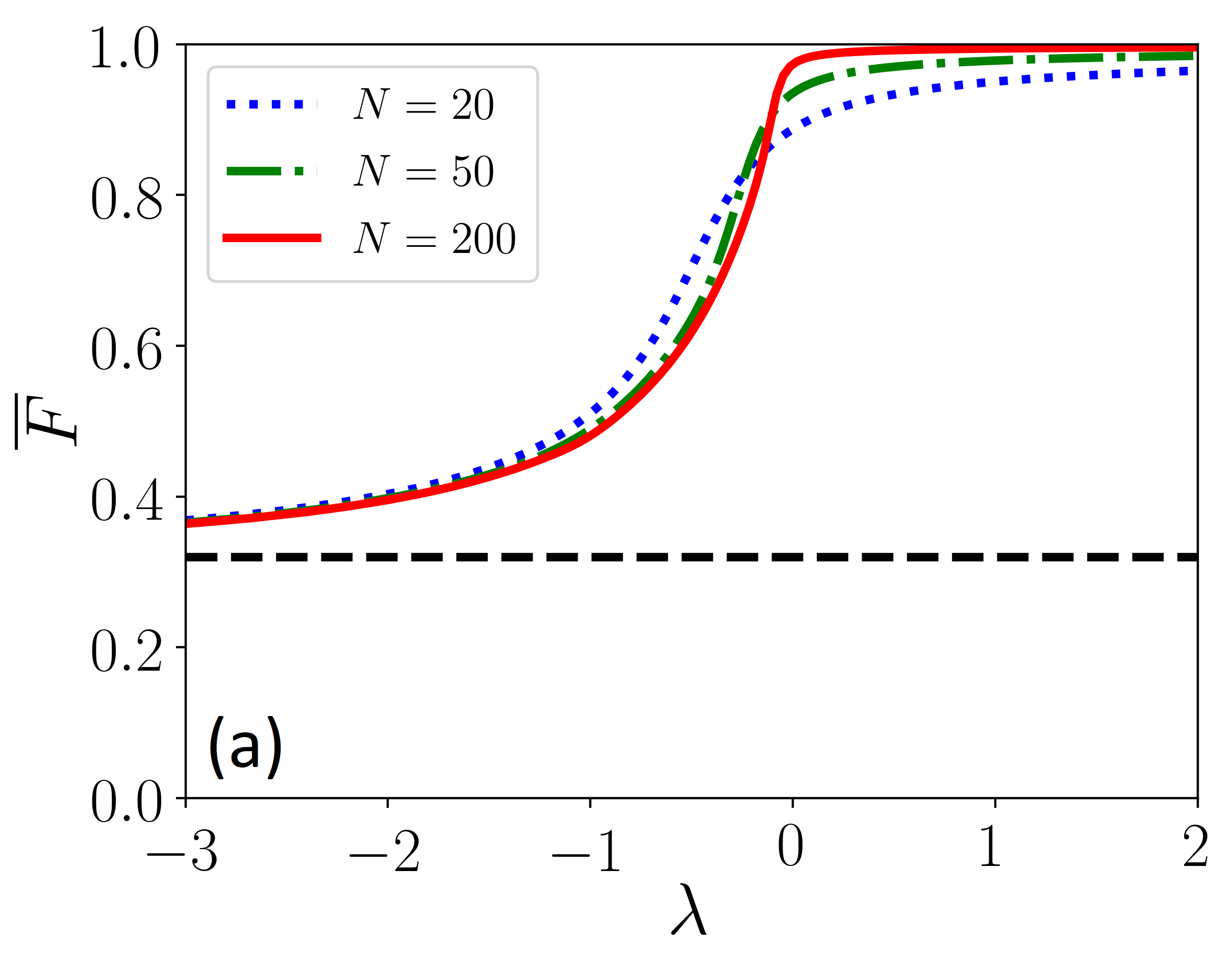} 
        \includegraphics[height=4.5cm,width=6cm]{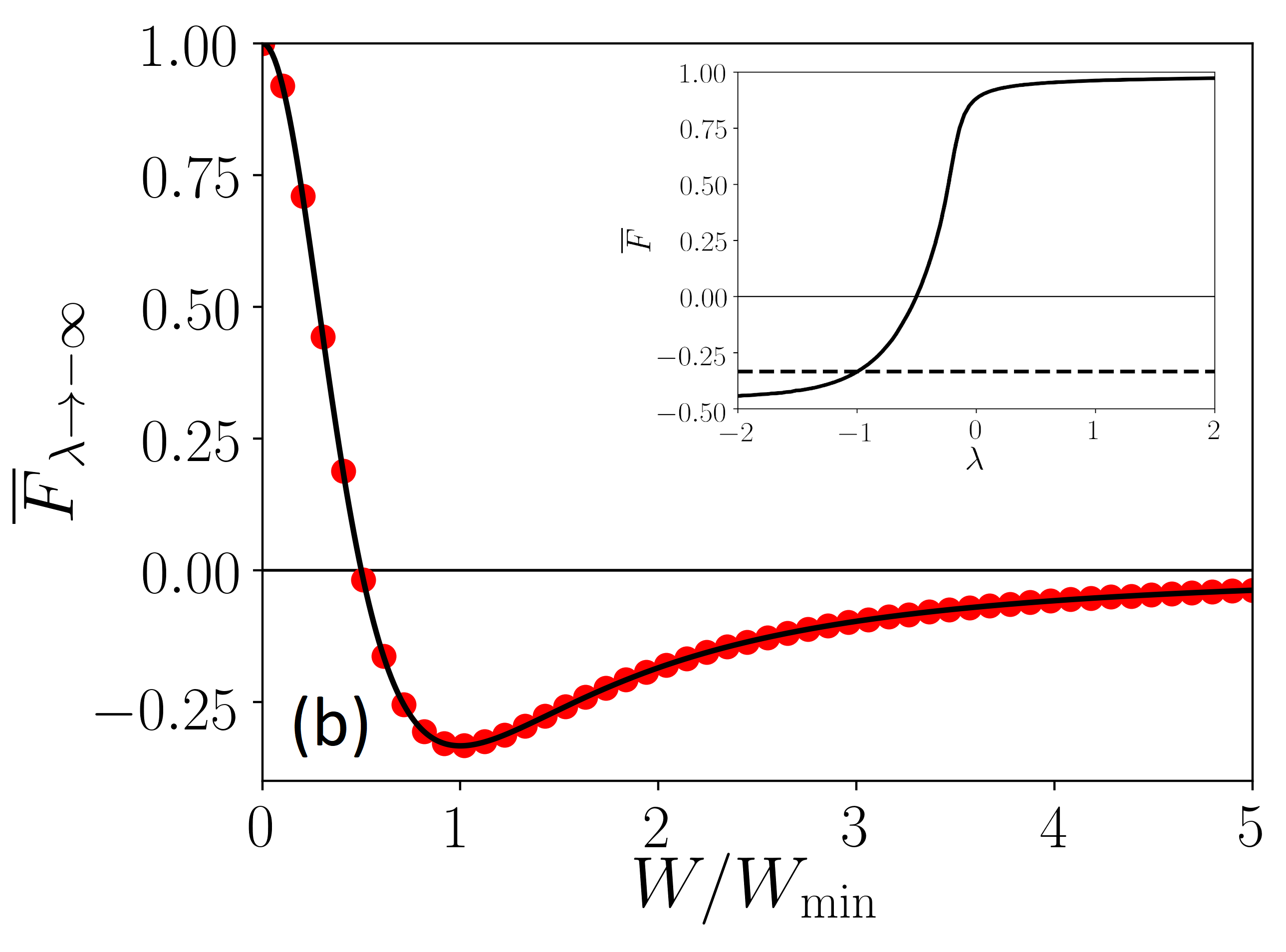}
        \includegraphics[height=4.5cm,width=5.5cm]{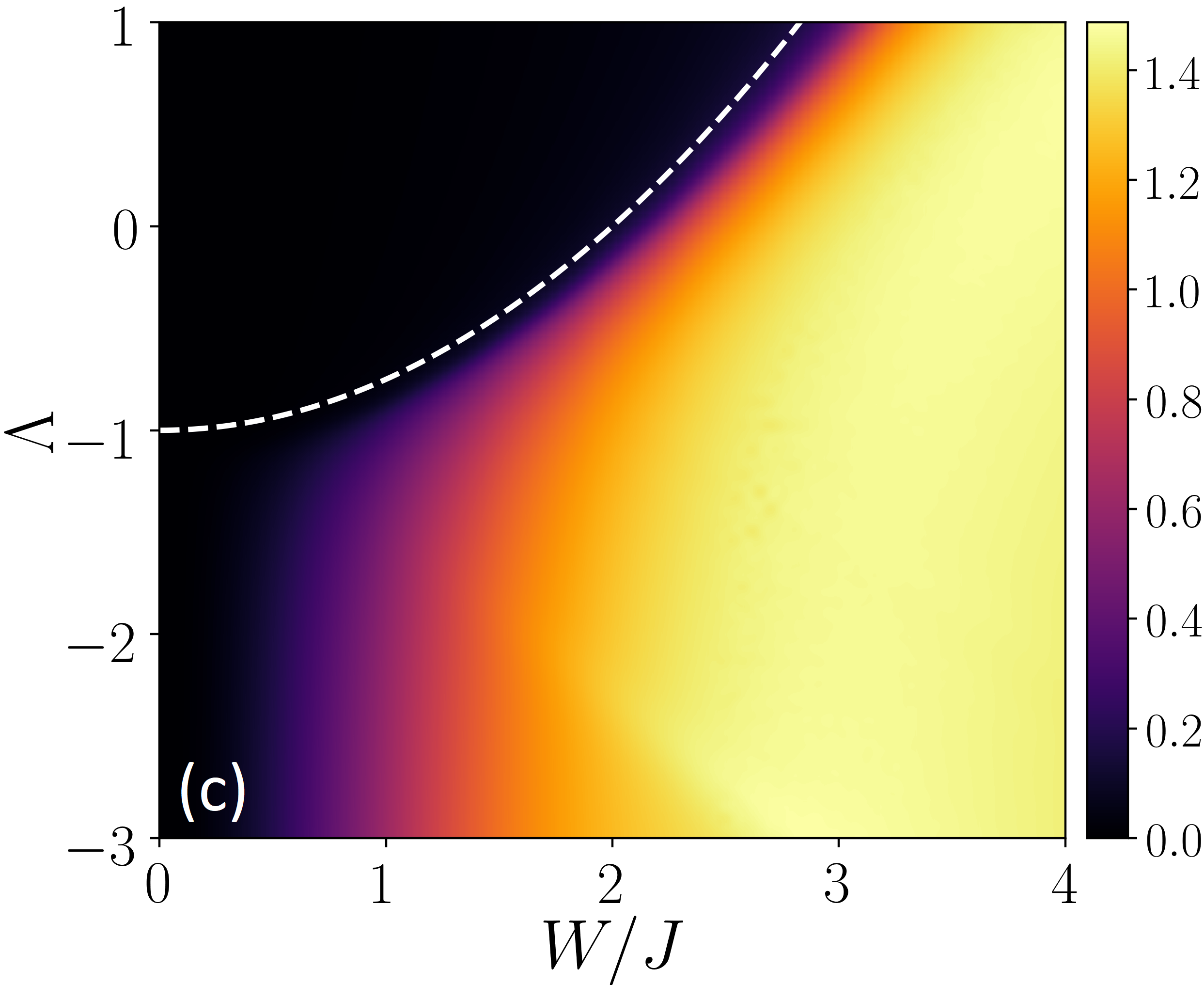} 
        \caption{The long-time average of the qubit OTOC, $\overline{F}$, is displayed as a function of $\lambda = (\Lambda - \Lambda_c)/\vert \Lambda_c \vert$ and $W$.  In (a) $\overline{F}$ is plotted as a function of $\lambda$ for different system sizes: $N = 20$ (blue, dotted), $N = 50$ (green, dot-dashed) and $N=200$ (red, solid).  $\overline{F}$ is shown to be close to unity for $\lambda > 0$ (normal phase) and decreases at the critical point asymptotically approaching some value (dashed horizontal black line) as $\lambda \to -\infty$ (symmetry broken phase). In (b) we plot this asymptotic value as a function of $W$ for $N = 200$. The black curve is the theoretical value given in Eq.\ \eqref{eq:asympF} and the red dots are determined numerically showing excellent agreement.  Inset: $\overline{F}$ as a function of $\lambda$ for $W = W_{\mathrm{min}} = 2 \sqrt{2} J^a$ ($\Lambda_c = 1$) showing that $\overline{F}$ can be negative in the symmetry broken phase.  In (c) we construct a phase diagram given by a density plot of half of the out-of-time-ordered commutator $C(t)/2 = 1-\mathrm{Re}[\overline{F}]$ versus $\Lambda$ and $W$.  The white dashed curve is the critical line given by Eq.\ \eqref{eq:critlam}.  In all images $J^a = 1$.}
\label{fig:otocF}
\end{figure*}

In this paper, we use the OTOC of the qubit as a probe of the QPT in $\hat{H}$.  Unless specified otherwise, we choose the operators in Eq.\ \eqref{eq:otoc} to be: $\hat{A} = \hat{B} = \hat{\sigma}_x$, and $\langle ... \rangle$ is the expectation value with respect to the ground state of $\hat{H}$, $\vert \psi_0 \rangle$.  In Ref.\ \cite{heyl18} it was shown that the long-time average of Eq.\ \eqref{eq:otoc}, 
\begin{equation}
\overline{F} = \lim \limits_{T \to \infty} \frac{1}{T} \int_0^T \mathrm{d}t\;F(t),
\end{equation}
can be used to identify QPTs and this is the case here.  In Fig.\ \ref{fig:otoccol} (a) and (b) we show samples of the dynamics of $\mathrm{Re}[F(t)]$ (blue) and $\mathrm{Im}[F(t)]$ (red) in the normal and symmetry broken phases, respectively.  The imaginary part of the OTOC oscillates around zero in both phases, and is thus insensitive to the QPT. The real part, however, is sensitive to the QPT and oscillates around unity in the normal phase and less than unity in the symmetry broken phase.  For this reason, henceforth $\overline{F}$ will refer to $\mathrm{Re}[\overline{F}]$.

Figure\ \ref{fig:otocF}(a) shows that $\overline{F}$ of the qubit does a good job at diagnosing the QPT where a steep decrease is seen around $\lambda = 0$ where $\lambda \equiv (\Lambda - \Lambda_c)/\vert \Lambda_c \vert$ is the reduced driving parameter.  The steepness of the curves increases with system size, creating a kink at $\lambda = 0$ in the thermodynamic limit.  The asymptotic behaviour of all three curves is the same, however, and can be explained by taking the limits $\Lambda \to \pm \infty$ in Eq.\ \eqref{eq:ham}.  In the limit $\Lambda \to +\infty$, $\langle \hat{S}_z \rangle = 0$, so the effective Hamiltonian in the normal phase becomes $\hat{H} = - N J^a \hat{\sigma}_x$.  Since $[\hat{H}, \hat{\sigma}_x] = 0$ in this phase the OTOC is time independent and takes the value $F = \langle \left ( \hat{\sigma}_x \right )^{4} \rangle = 1$.  In the opposite limit of $\Lambda \to - \infty$, $\langle \hat{S}_z \rangle = \pm N/2$, so the effective Hamiltonian in the symmetry broken phase is $\hat{H} = N \left ( - J^a \hat{\sigma}_x \pm W \hat{\sigma}_z \right )$ resulting in an analytic, but cumbersome expression for $F(t)$.  However, $\overline{F}$ has the simple expression  

\begin{equation}
\overline{F}_{\lambda \to - \infty} = 
\lim_{\lambda \to -
\infty} \overline{F} = \frac{8 J^{a2} \left (2 J^{a2} - W^2 \right )}{\left (4 J^{a2} + W^2 \right )^2} \, .
\label{eq:asympF}
\end{equation}
In Fig.\ \ref{fig:otocF}(b) we compare Eq.\ \eqref{eq:asympF}  (black curve) as a function of $W$ against numerical results (red circles) for $J^a = 1$, $\Lambda = - 500$, and $N = 200$ and find excellent agreement.  The most notable feature in the image is that $\overline{F}$ can be negative which is a departure from the case where the qubit is absent and one of the collective boson operators is used in Eq.\ \eqref{eq:otoc} as $\hat{A}$ and $\hat{B}$ instead.  We find that in the limit $\Lambda \to -\infty$, $\overline{F}$ reaches a minimum of -1/3 at $W_{\mathrm{min}} =2 \sqrt{2} J^a$.  Thus, if the location of the critical point is unknown, then when $W$ is properly tuned to be near $W_{\mathrm{min}}$ one only needs to know that $\overline{F}$ has changed sign to locate the critical region rather than its exact value, making locating the QPT easier.

In fact, the QPT can be located for almost all values of $\Lambda$ and $W$ using $\overline{F}$, or in this case $\overline{C}/2 = 1 - \overline{F}$, as shown in Fig. \ref{fig:otocF}(c).  We find that the dark (normal) and bright (symmetry broken) phases found from $\overline{C}$ agree well with $\Lambda_{c}$  as given by Eq.\ \eqref{eq:critlam} and plotted as the white dashed curve.  The dark strip along the $W = 0$ axis does not mean there is no QPT at $\Lambda_c = -1$. Instead it means that the qubit fails to detect it because it is no longer coupled to the $N$ bosons. 

 In strongly chaotic systems it is expected that $F(t \to \infty) = 0$ or $C(t\to\infty)/2 = 1$ \cite{roberts15} because the system equilibrates to a point where there are no more correlations between operators due to information scrambling.  This reasoning allowed the authors of Ref. \cite{buijsman17} to identify regions of chaos in the anisotropic Dicke model in a plot similar to Fig. \ref{fig:otocF}(c).  In our case, the backaction of the impurity qubit on the $N$ bosons causes the mean-field (classical) version of $\hat{H}$ to produce chaotic dynamics in the symmetry broken phase \cite{mumford14a}, therefore we should also expect $F(t \to \infty) = 0$ and $C(t\to\infty)/2 = 1$ in that phase.  However, looking at Fig.\ \ref{fig:otoccol}(b) we don't see signs of the expected equilibration around zero in the symmetry broken phase.  In the next section we will delve into why this is the case and when we should expect to get the chaotic result for the OTOC.


\section{Effects of Backaction}
\label{sec:equil}

To gain some insight into how the qubit can be involved in equilibration, we follow \cite{dag19} and use the eigenstates of $\hat{H}$ to form the completeness relation $\sum_\gamma \vert \psi_\gamma \rangle \langle \psi_\gamma \vert = \mathds{1}$. Inserting this into Eq.\ \eqref{eq:otoc} the OTOC takes the general form
\begin{equation}
F(t)=\sum_{\alpha, \gamma, \gamma^\prime, \beta} c_\alpha^* b_\beta \mathrm{e}^{-\mathrm{i} \left (E_\beta - E_\alpha + E_\gamma - E_{\gamma^\prime} \right )t} B^\dagger_{\alpha \gamma} A_{\gamma \gamma^\prime}^\dagger B_{\gamma^\prime \beta}\; .
\label{eq:analotoc1}
\end{equation}
Here $\langle \psi_\alpha \vert \hat{B} \vert \psi_\gamma \rangle = B_{\alpha \gamma}$, $\langle \psi_\alpha \vert \psi (0) \rangle = c_\alpha$, $\langle \psi_\beta \vert \hat{A} \vert \psi (0) \rangle = b_\beta$, and $\vert \psi(0)\rangle$ is the general state used in Eq.\ \eqref{eq:otoc}, where in our case we put $\vert \psi (0) \rangle = \vert \psi_0 \rangle$, so that $c_\alpha = \delta_{0 \alpha}$ and $b_\beta = A_{\beta 0}$.  Keeping in mind that $\hat{B} = \hat{A} = \hat{\sigma}_x$, Eq.\ \eqref{eq:analotoc1} simplifies to

\begin{equation}
F(t)=\sum_{\gamma, \gamma^\prime, \beta} e^{\mathrm{i} \left (E_\beta - E_0 + E_\gamma - E_{\gamma^\prime} \right )t} \sigma_{0 \gamma} \sigma_{\gamma \gamma^\prime} \sigma_{\gamma^\prime \beta} \sigma_{\beta 0}
\label{eq:analotoc2}
\end{equation}
where $\langle \psi_\alpha \vert \hat{\sigma}_x \vert \psi_\gamma \rangle = \sigma_{\alpha \gamma}$ and we have removed the $x$ label for convenience.  In Eq.\ \eqref{eq:analotoc2}, it becomes clear that in order for $F(t)$ to equilibrate, $\hat{\sigma}_x$ must couple many energy eigenstates together, so there will be many phase factors contributing to the dynamics resulting in phase decoherence.  In the normal phase the ground state is approximately a product state of the form $\vert \psi_0 \rangle \approx \vert \psi_0 \rangle_B \otimes \left ( \vert + \rangle _Q + \vert - \rangle_Q \right )/\sqrt{2}$, so it is approximately an eigenstate of $\hat{\sigma}_x$ with eigenvalue $+1$.  Therefore, in the normal phase $\sigma_{00} \sim \mathcal{O}(1)$ and for all other matrix elements $\sigma_{\gamma \gamma^\prime} \ll 1$.  This is exactly what we see in Fig.\ \ref{fig:otoccol}(a) where $F(t)$ is very close to unity at all times and the oscillations are due to the finite size of the system.  In the symmetry broken phase it is expected that fluctuations between matrix elements of $\hat{\sigma}_x$ will be large, so many states get coupled together.  For this reason, there will be many small contributions to $F(t)$ leading to $F(t\to\infty) = 0$.  However, this is \textit{not} what we see in Fig.\ \ref{fig:otoccol}(b) where $F(t)$ has large fluctuations over the time interval shown [we have checked for times $\mathcal{O}(10^{5})$ longer and still find no equilibration for the parameter values in Fig.\ \ref{fig:otoccol}].  We expect that although $\hat{\sigma}_x$ does couple more eigenstates together in the symmetry broken phase, due to the small values of $W$ used thus far, the qubit does not have a large effect on the $N$ boson system.  Therefore, as $W$ increases $\hat{\sigma}_x$ should couple more eigenstates together and $F(t)$ will reach a quasi-equilibrium state at long times.  We say quasi-equilibrium because there will still be revivals, however, their spacing in time should increase as the number of states contributing to $F(t)$ increases.

\begin{figure*}[t!]
        \includegraphics[height=4.5cm,width=5.5cm]{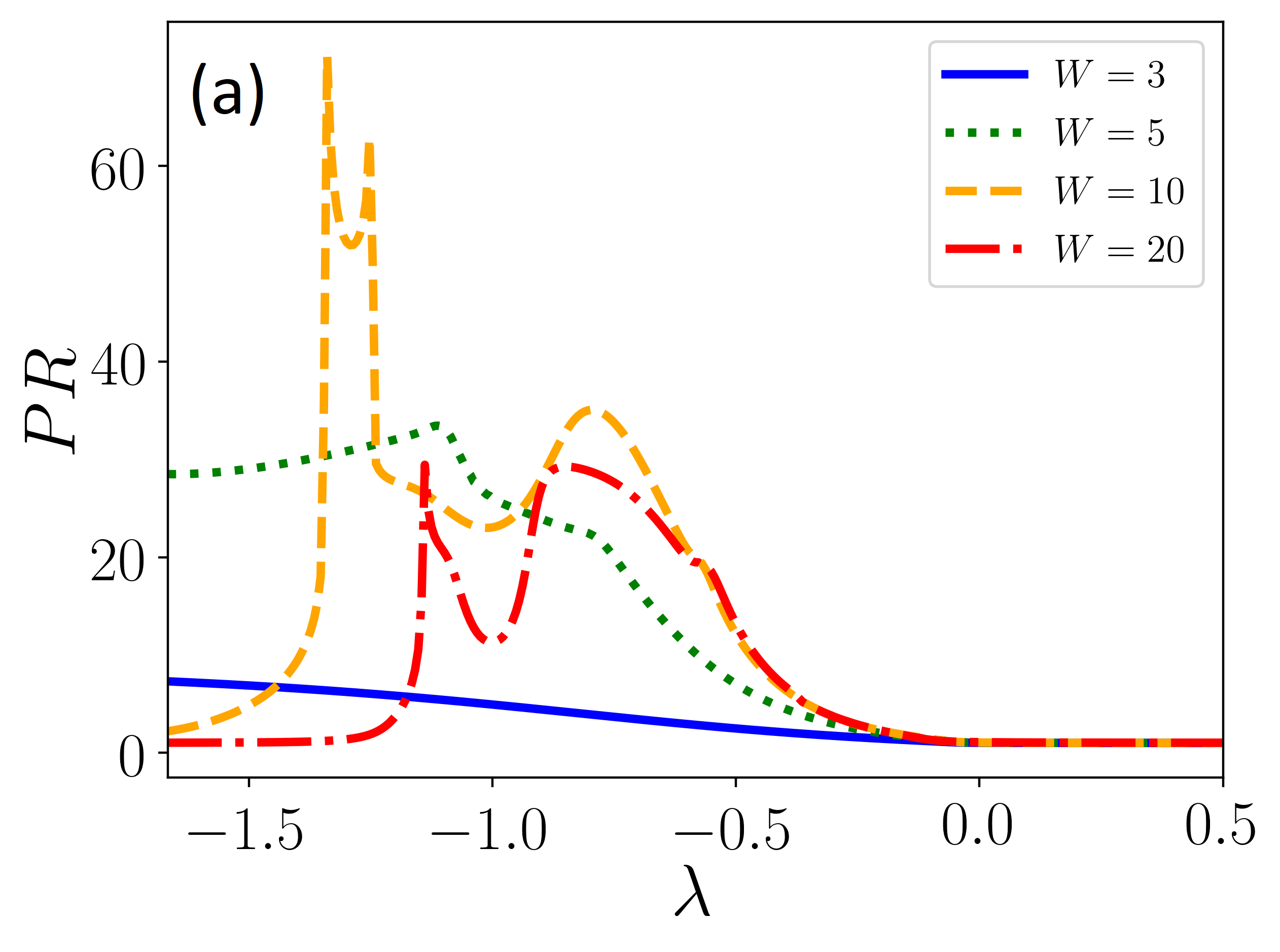} 
        \includegraphics[height=4.5cm,width=6.5cm]{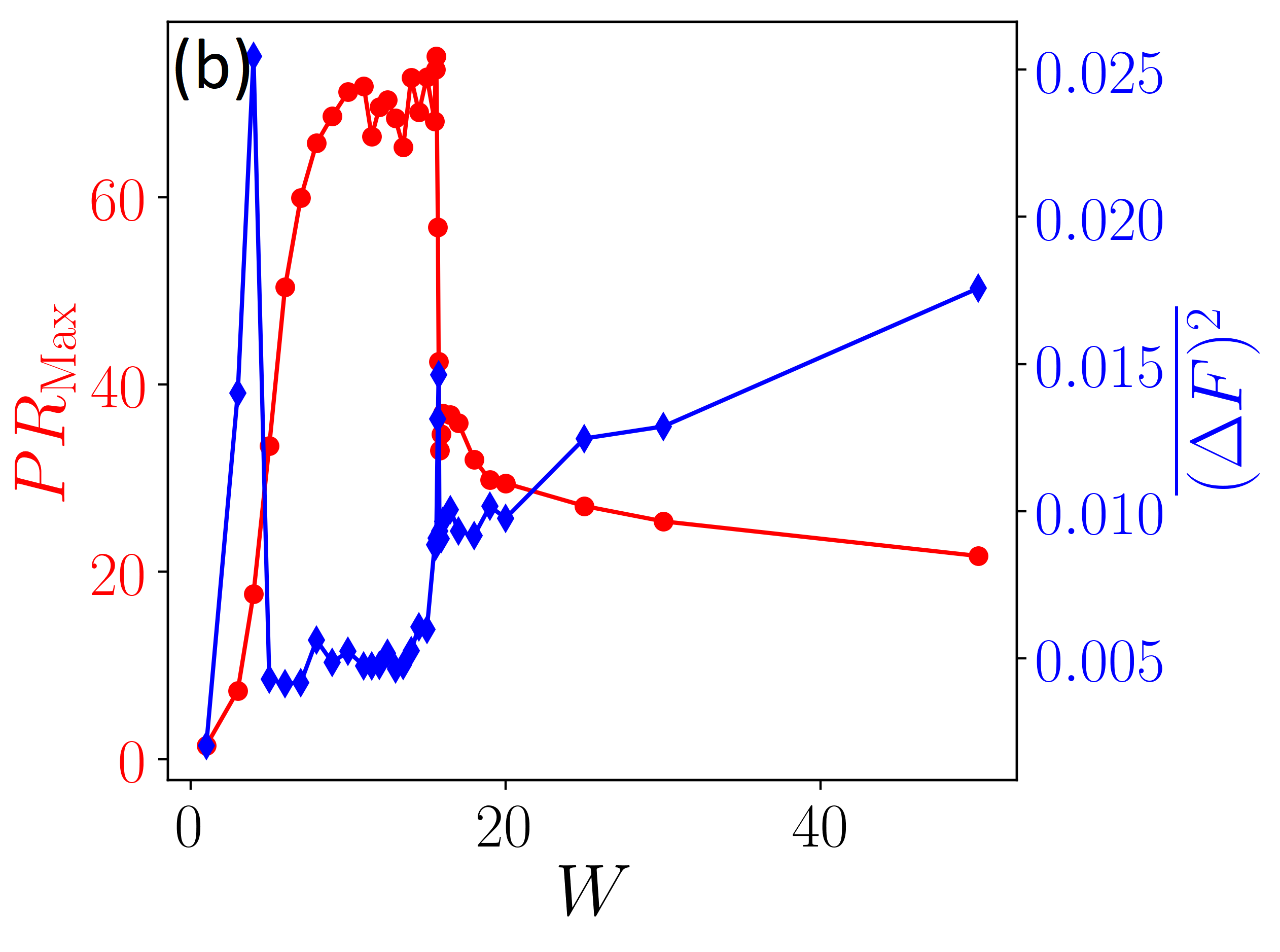}
        \includegraphics[height=4.5cm,width=5.5cm]{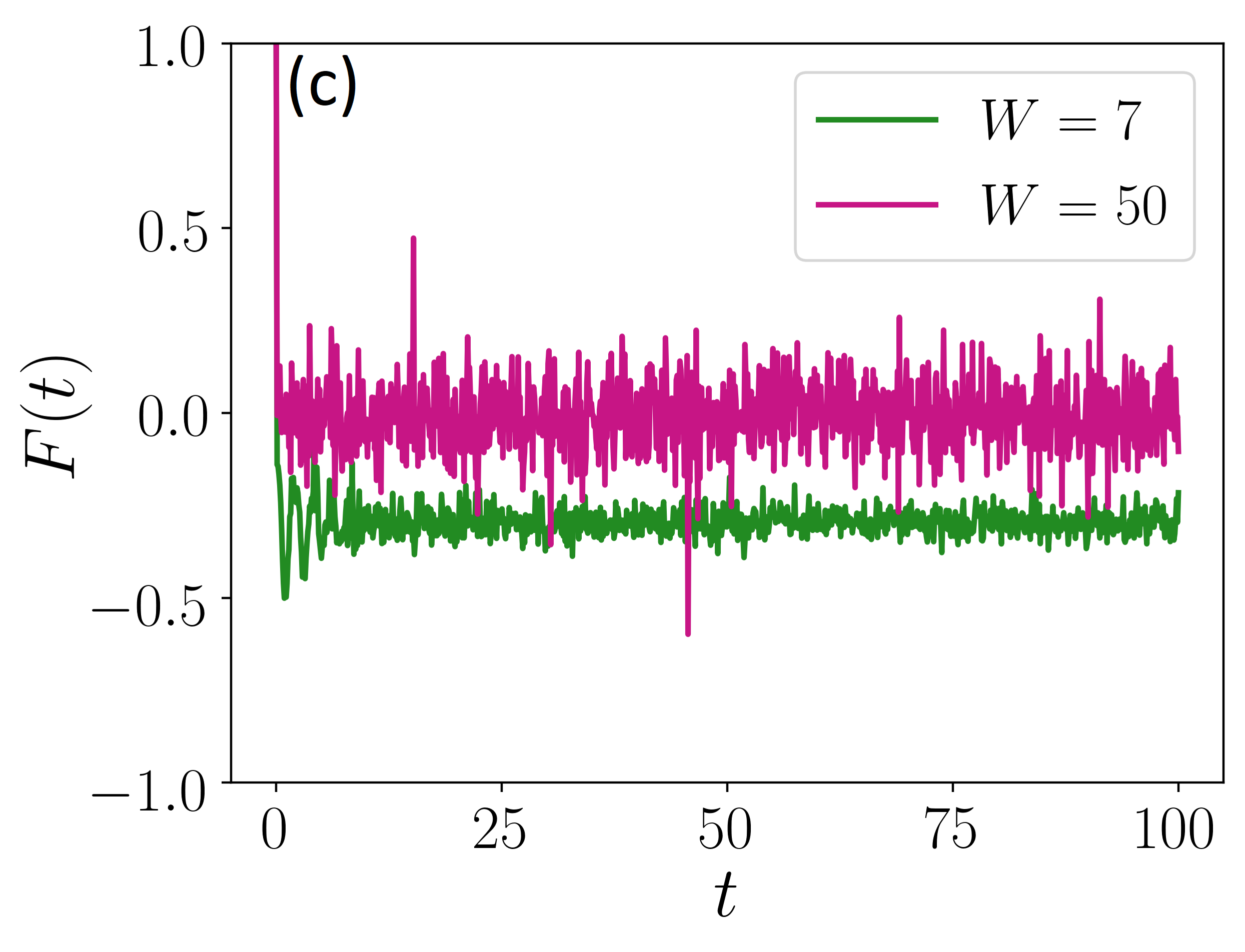} 
        \caption{The participation ratio (PR), the long-time average of the OTOC variance $\overline{(\Delta F)^2}$, and the equilibration of $F(t)$ are shown for different values of $\lambda$ and $W$.  Image (a) shows Eq.\ \eqref{eq:PRQ} as a function of $\lambda$ for different values of $W$.  The curves are nonmonotonic and reach a maximum at different values of $\lambda$ for each value of $W$ which we denote as $PR_{\mathrm{Max}}$.  Image (b) has $PR_{\mathrm{Max}}$ (red circles) as a function of $W$ plotted alongside $\overline{(\Delta F)^2}$ (blue diamonds).  The peak values of $PR_{\mathrm{Max}}$ coincide with a drop in $\overline{(\Delta F)^2}$ confirming the idea that the more energy eigenstates there are contributing to $F(t)$, the smaller the temporal fluctuations will be.  Image (c) has $F(t)$ for $W=7$ (green) and $W=50$ (purple) showing the dynamics get closer to a quasiequilibrium state at long times for values of $W$ and $\lambda$ where $PR_{\mathrm{Max}}$ is a maximum.  For all of the images $N=1000$ and $J^a = 1$.}
\label{fig:PR}
\end{figure*}

To help quantify the increased eigenstate coupling due to both the QPT and increased $W$ we use the participation ratio (PR). This determines how spread a reference state is in a particular basis and is defined as
\begin{equation}
PR = \left ( \sum_{n} \vert \langle n \vert a \rangle \vert^4 \right )^{-1}\;,
\end{equation} 
where $\vert a \rangle$ is the reference state and $\left \{ \vert n \rangle \right \}$ is the set of basis states belonging to the Hilbert space of interest.  The inverse of the PR has been shown to be related to OTOCs following a quench when the operators in Eq.\ \eqref{eq:otoc} are the projection operators of the eigenstates of the prequench Hamiltonian \cite{borgonovi19}, so it is a relevant quantity to investigate here.  The minimum value the PR can take is unity when the reference state is entirely localized in the Hilbert space $\vert a \rangle = \vert n^\prime \rangle$.  When the reference state is spread equally throughout the Hilbert space, $\vert a \rangle = (1/\sqrt{\mathcal{N}}) \sum_{n^\prime} \vert n^\prime \rangle$, the PR reaches its maximum value of $\mathcal{N}$ where $\mathcal{N}$ is the size of the Hilbert space.  In our case, the reference state is $\hat{\sigma}_x \vert \psi_0 \rangle$ ($\vert \psi_0 \rangle$ is the ground state of $\hat{H}$) and the basis of interest is formed by the energy eigenstates of $\hat{H}$, so 
\begin{equation}
PR =  \left ( \sum_n \vert \langle \psi_n \vert \hat{\sigma}_x \vert \psi_0 \rangle \vert^4\right )^{-1} \, .
\label{eq:PRQ}
\end{equation}
In the normal phase $\vert \psi_0 \rangle$ is almost an eigenstate of $\hat{\sigma}_x$ for large enough $N$, so $PR \approx 1$.  In the symmetry broken phase $\hat{\sigma}_x \vert \psi_0 \rangle$ is spread over many energy eigenstates, so the PR should increase.  In Fig.\ \ref{fig:PR}(a) we plot the PR as a function of $\lambda$ for different values of $W$ and we observe the expected behaviour: a drastic increase in the PR toward the symmetry broken phase.  Although the PR curves are complicated functions of $\lambda$,  for each value of $W$ we are most interested in the value of $\lambda$ where the PR is a maximum because we expect $F(t)$ to be closest to equilibrium at long times there.  We denote these maximum values as $PR_{\mathrm{Max}}$ and plot them as a function of $W$ in Fig.\ \ref{fig:PR}(b) alongside the long-time average of the variance of the OTOC, $\overline{(\Delta F)^2} = \lim_{T \to \infty} 1/T \int_0^T dt \left ( \overline{F} - F(t) \right )^2$.  We find that the peak in $PR_{\mathrm{Max}}$ coincides almost perfectly with a drop in the variance of the OTOC confirming that the more energy eigenstates participating in the sum in Eq.\ \eqref{eq:analotoc2} the closer the system is to an equilibrium state.  To emphasize this point further we plot $F(t)$ in Fig.\ \ref{fig:PR}(c) for values of $W$ corresponding to large ($W=7$) and small ($W=50$) $PR_{\mathrm{Max}}$ which shows the temporal fluctuations are smaller for larger $PR_{\mathrm{Max}}$.

An important feature demonstrated by Fig.\ \ref{fig:PR}(c) is that $F(t)$ is closest to equilibration, not at $F=0$, but at some value $F<0$.  We could increase $W$ further to try to obtain $F(t\to\infty) = 0$ as $W\to\infty$, similar to how $\overline{F} = 0$ as $W\to\infty$ in Fig.\  \ref{fig:otocF}(b).  However, this does not work because in this limit $\hat{H} \rightarrow \hat{H}_I =  W \hat{S}_z \hat{\sigma}_z$ and $\vert \psi_0 \rangle \rightarrow \left (\vert N/2 \rangle \vert - \rangle + \vert -N/2 \rangle \vert + \rangle \right )/\sqrt{2}$ which is a Schr\"{o}dinger cat state made up of a superposition of the two extreme bosonic $S_{z}$ states (and in each case they are perfectly anticorrelated with the qubit which is in the opposite spin state).  This gives $F(t) = \cos (WNt/2)$ which does indeed have an infinite time average of zero, however, it also has the largest possible fluctuations in time.  Therefore, with regards to the OTO commutator, the best the system can do is $C(t\to\infty)/2 > 1$ instead of the fully scrambled and fully chaotic result of $C(t\to\infty)/2 = 1$. We therefore conclude that the qubit OTOC shows signs of weak chaos and exhibits weak information scrambling. This is in agreement with our previous work \cite{mumford14a} where the level spacing statistics of $\hat{H}$ for $\Lambda=0$ (i.e.\ in the absence of boson self-interaction $U=0$) were analyzed and, although we found evidence of level repulsion in the symmetry broken phase,  the statistics did not obey a fully-fledged Wigner-Dyson distribution suggesting weak but not full chaos.   

\section{Finite Size Scaling}

\begin{figure}[t!]
        \includegraphics[height=3.25cm]{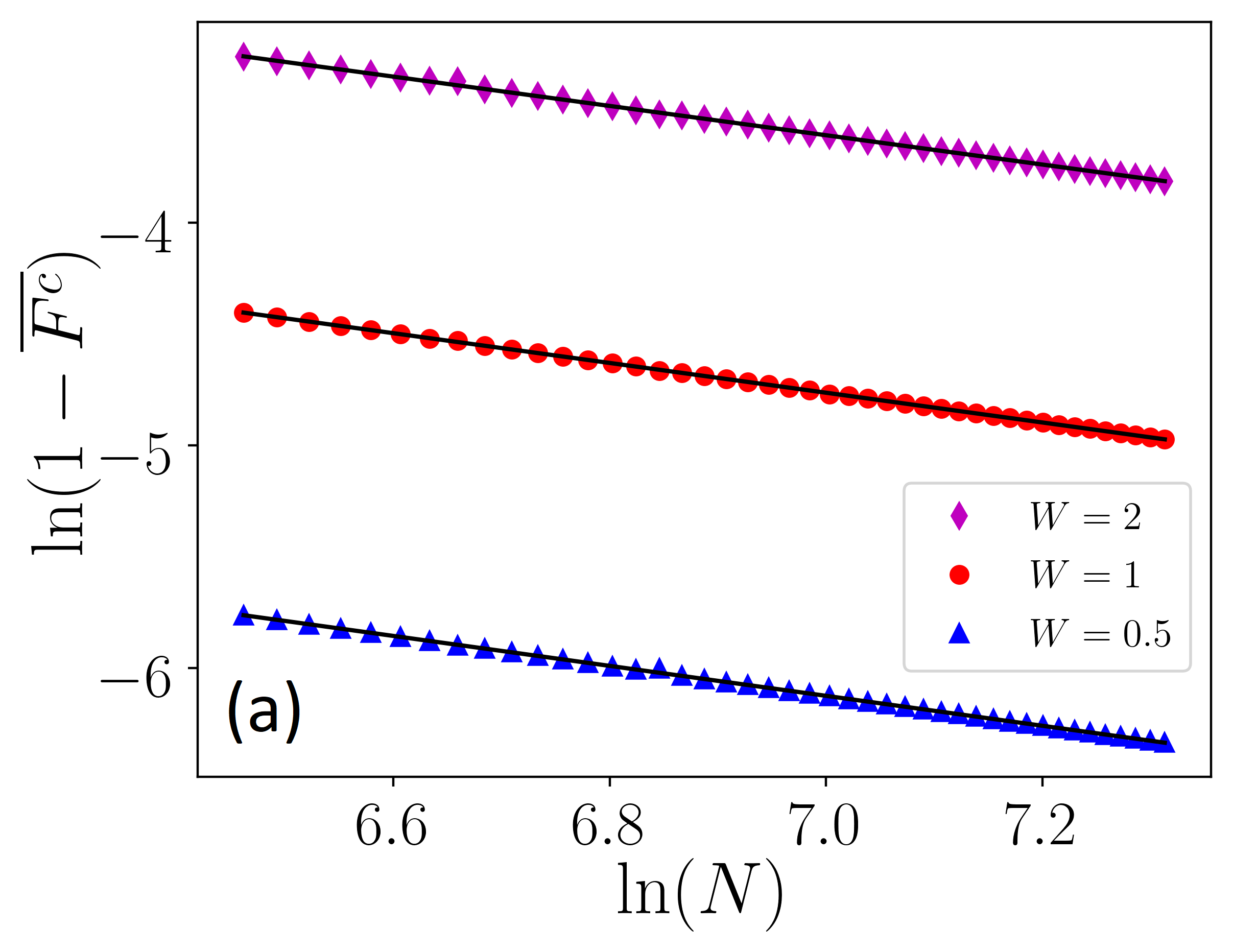} 
        \includegraphics[height=3.25cm]{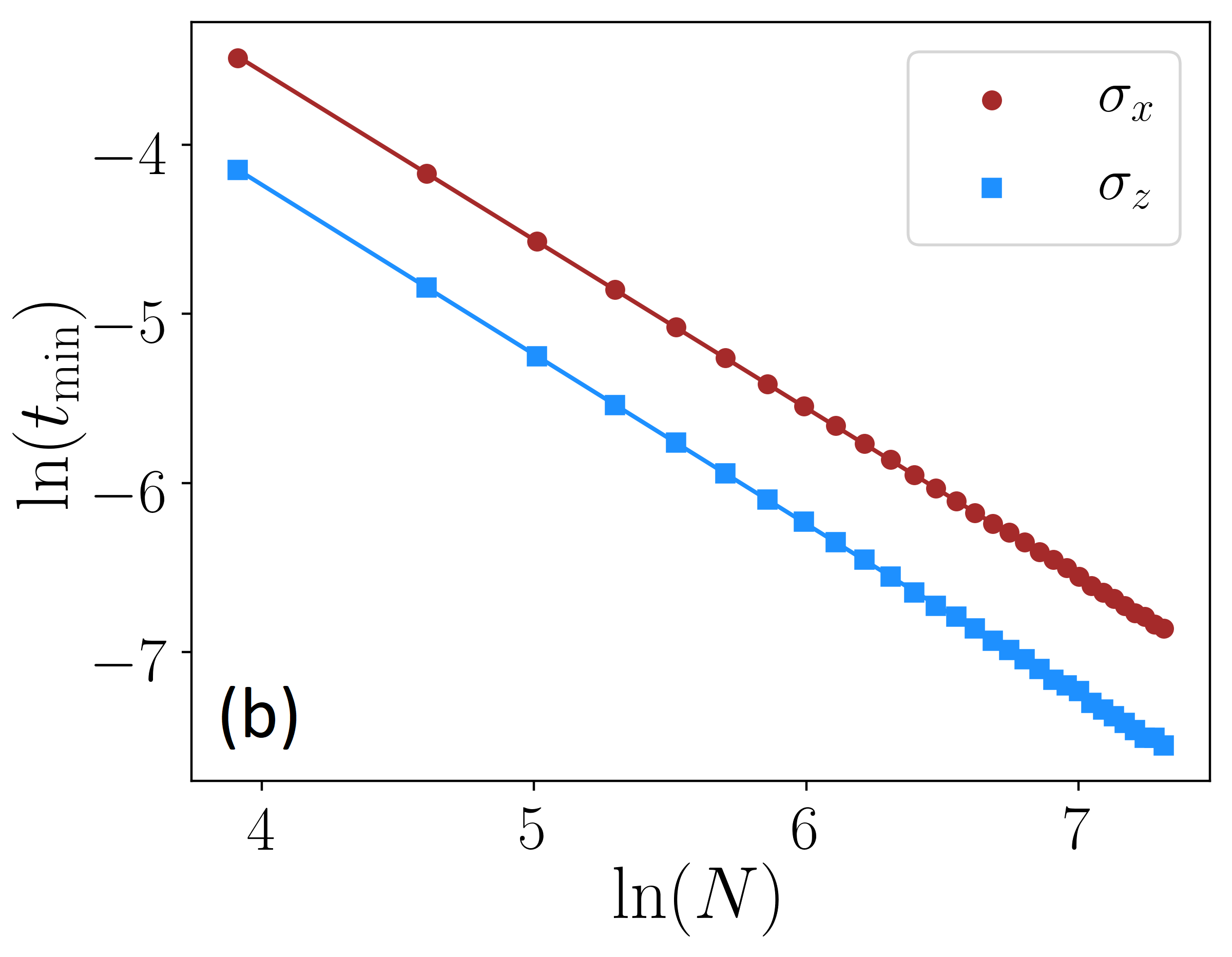} 
        \caption{Finite-size scaling of the OTOC.  Image (a) shows a Log-Log plot of $1-\overline{F^c}$ versus $N$ for different values of $W$ where $\overline{F^c}$ is the long time average of the OTOC at $\Lambda = \Lambda_c$.  The slopes are $b=0.658$ ($W=2$), $b=0.668$ ($W=1$) and $b=0.672$ ($W=0.5$).  Image (b) shows a Log-Log plot of the $t_{\mathrm{min}}$ versus $N$ for $\hat{A} = \hat{B} = \hat{\sigma}_x$ (red circles) and $\hat{A} = \hat{B} = \hat{\sigma}_z$ (blue squares) where $t_{\mathrm{min}}$ is the time at which the first minimum of $F(t)$ occurs.  The slopes are $d=-0.995$ ($\hat{\sigma}_x$) and $d=-1.002$ ($\hat{\sigma}_z$).  The system size range is $50 \leq N \leq 1500$.}
\label{fig:FSS}
\end{figure}

In the thermodynamic limit we expect $\overline{F}$ to change suddenly at the critical point: being exactly unity when $\Lambda > \Lambda_c$ and less than unity when $\Lambda < \Lambda_c$.  Determining how $\overline{F}$ scales with system size helps characterize the QPT, allowing for a more rigorous comparison with QPTs in other systems.  To this end, Fig.\ \ref{fig:FSS}(a) shows a Log-Log plot of $1- \overline{F^c}$  as a function of $N$ for different values of $W$, where $\overline{F^{c}}$ is the long time average of $F(t)$ at $\Lambda=\Lambda_{c}$.   From this, we determine that for larger system sizes, $\overline{F^c}$ approaches unity.  The linear dependence indicates that $1- \overline{F^c} \sim N^{-b}$ where the slopes give exponents: $b \approx 0.658$ for $W = 2.0$, $b \approx 0.668$ for $W=1.0$, and $b \approx 0.672$ for $W=0.5$, suggesting an exponent of $b = 2/3$. This value can be understood by considering Eq.\ \eqref{eq:analotoc2}, and in particular noting that at the quantum critical point $\Lambda_c$ (and above it) the largest contribution to the sum is the term where $\beta = \gamma = \gamma^\prime = 0$ giving: 
\begin{equation}
\overline{F^c} = \left ( \sigma_{00} \right )^4 \;.
\label{eq:critotoc}
\end{equation}
We have previously shown that for $J^a = 1$, in the ground state $\left \langle \hat{\sigma}_x \right \rangle$ has similar dependence as $2\braket{\hat{S}_x } /N$ \cite{mumford14a,mumford14b}, while at the critical point it has been shown that \cite{dusuel04}
\begin{equation}
2  \braket{\hat{S}_x} /N \sim 1 + 1/N + a_x N^{-2/3}
\end{equation}
where $a_x$ is an $N$ independent constant.  For large $N$, $1-2 \braket{\hat{S}_x} /N$ therefore scales as $N^{-2/3}$, and so we expect $1-\left \langle \hat{\sigma}_x \right \rangle$ to scale in a similar fashion.  Substituting this scaling into Eq.\ \eqref{eq:critotoc}, to leading order we get $1-\overline{F^c} \sim N^{-2/3}$ agreeing with the numerical results.

Recently, a finite size scaling theory of OTOCs has been proposed in Ref. \cite{wei19} where for this particular system the scaling of the OTOC function at the critical point is expected to take the general form,
\begin{equation}
F(t) = f( N^{-z} t)\;,
\end{equation}
where $z$ is the dynamical critical exponent and $f(x)$ is some function of $x$.  The Hamiltonian used in Ref. \cite{wei19} is $\hat{H}_B$ (no qubit) with OTOC operators $\hat{A} = \hat{B} = \hat{S}_z$ in Eq.\ \eqref{eq:otoc}.  They extracted the value $z=1/3$ from their OTOC numerically via a Log-Log plot of $t_{\mathrm{min}}$ versus $N$, where $t_{\mathrm{min}}$ is the time at which the first minimum of $F(t)$ occurs.  In the system we consider here, we confirm that the value of $z = 1/3$ persists for the OTOC operator  based on $\hat{S}_z$ even in the presence of the qubit probe. However, when considering the OTOC for the qubit (via $\hat{\sigma}_x$ or $\hat{\sigma}_z$, for example) we encounter different behaviour. Fig.\ \ref{fig:FSS}(b) shows a linear dependence in Log-Log plots of $t_{\mathrm{min}}$ versus $N$, and thus $t_{\mathrm{min}} \sim N^d$. From the slopes we extract the exponents $d =-0.995$ for $\hat{\sigma}_x$ and $d = -1.002$ for $\hat{\sigma}_z$, suggesting a scaling exponent of $d=-1$.  

To explain the discrepancy between $z$ and $d$ we note that if we evaluate the qubit OTOC at $t_{\mathrm{min}}$ we should expect its early time dynamics to be dominated by the qubit term, $\hat{H}_Q$, in the Hamiltonian.  The factor of $N$ in $\hat{H}_Q$, which we inserted so $\Lambda_c \neq 0$ in the thermodynamic limit, only makes $\hat{H}_Q$ more dominant at early times and it is the source of the $d = -1$ exponent in $t_{\mathrm{min}} \sim N^d$.

%
%

\section{Excited state phase transitions}

\begin{figure}[t!]
        \includegraphics[height=3.25cm]{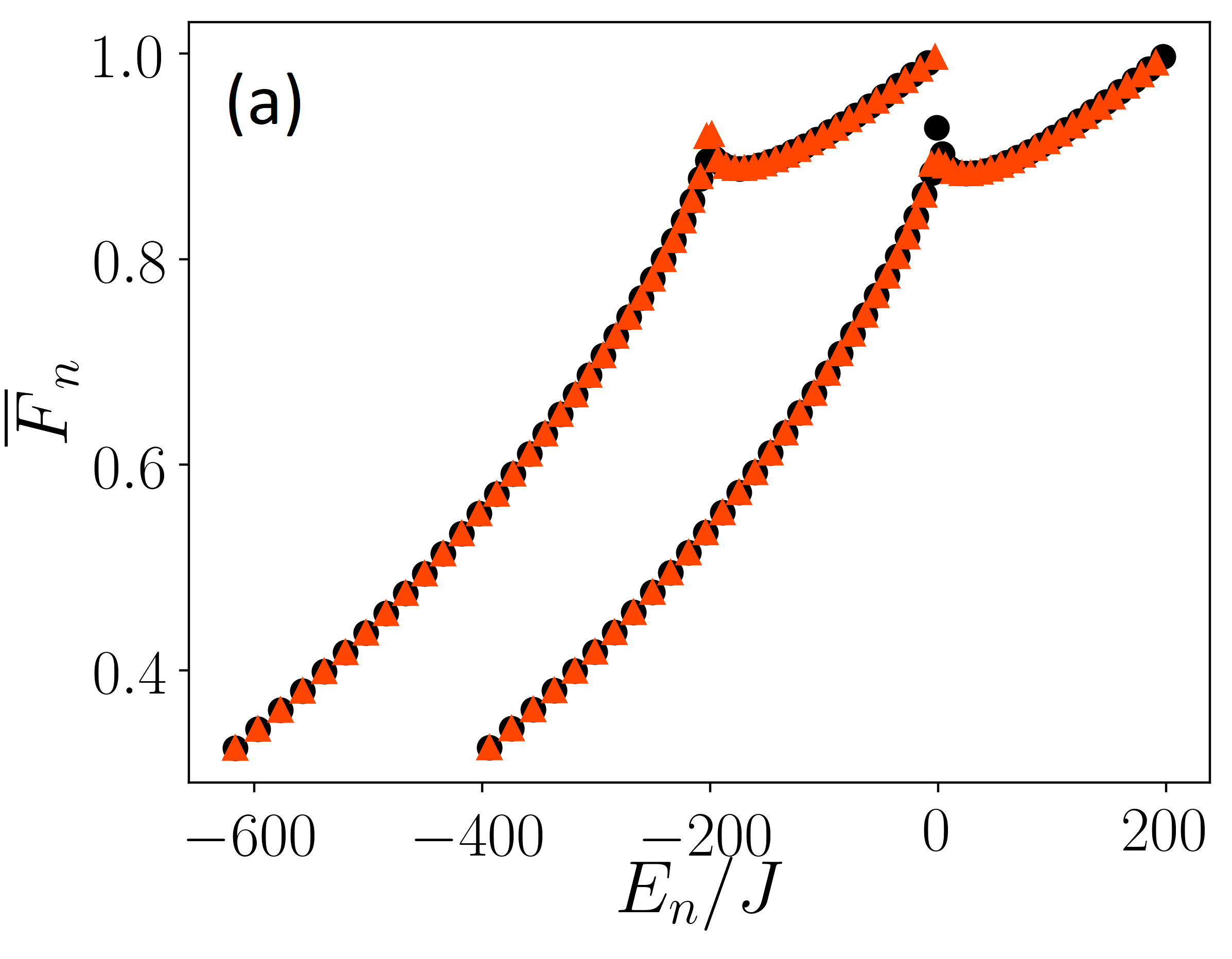} 
        \includegraphics[height=3.25cm]{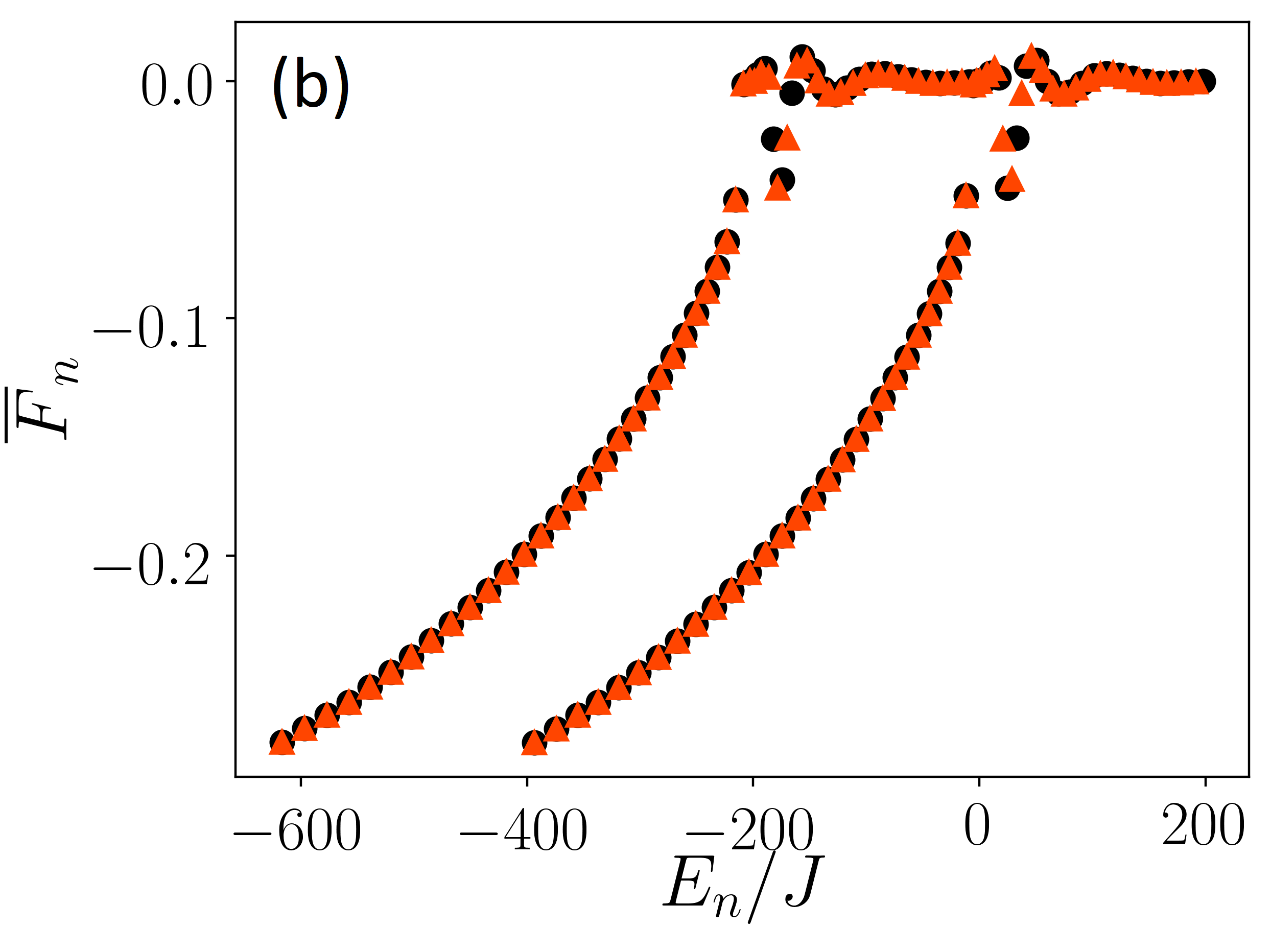} 
        \caption{Identifying ESPQTs using qubit OTOCs: $\overline{F}_n$, as defined in Eq.\ (\ref{eq:ESotoc}), plotted as a function of the energy eigenvalues $E_{n}$.  Images (a) and (b) are generated using $\hat{A}=\hat{B}=\hat{\sigma}_x$ and $\hat{A} = \hat{B} = \hat{\sigma}_z$ in Eq.\ \eqref{eq:ESotoc}, respectively.  The black circles and orange triangles represent even and odd parity states, respectively, and the parameter values are $W = J^a = 1$, $N = 100$ and $\Lambda = -10$ giving $E_c = -200$ and $E_c = 0$ depending on the spin state of the qubit which is used to evaluate the OTOC.}
\label{fig:ESotoc}
\end{figure}

In this section we explore some ramifications for OTOCs of the fact that for $\Lambda < \Lambda_c$ the excited energy eigenstates can undergo similar QPTs as the ground state. These are called excited state quantum phase transitions (ESQPTs).  An ESQPT manifests itself as a peak (that becomes a singularity in the thermodynamic limit) in the density of states at a critical energy $E_c$.  To help investigate ESQPTs we define the OTOC of the $n^{\mathrm{th}}$ energy eigenstate (of energy $E_{n}$) as, 
\begin{equation}
F_n(t) =  \bra{\psi_n}\hat{B}(t)^\dagger \hat{A}(0)^\dagger \hat{B}(t) \hat{A}(0) \ket{\psi_n} \, .
\label{eq:ESotoc}
\end{equation}
In Ref. \cite{wang18} it was shown that for $\hat{H}_B$ and $\hat{A} = \hat{B} = \hat{S}_z/N$, $\overline{F}_n$ is a good order parameter for the ESQPT where $\overline{F}_n = 0$ for $E_{n} \geq E_c$ and $\overline{F}_n \neq 0$ for $E_{n}<E_c$.  The critical energy $E_c$ is simply defined as the energy of the ground state in the normal phase.  An expression for $E_c$ can be found from the mean field version of Eq.\ (\ref{eq:ham}) giving $E_c = - N \left ( 1 \pm J^a \right )$ (in units of $J$) where the $\pm$ comes from the two states of the qubit (without the qubit $E_c = -N$).  In Fig.\ \ref{fig:ESotoc} we plot $\overline{F}_n$ as a function of the eigenenergy $E_n$ for $\hat{\sigma}_x$ and $\hat{\sigma}_z$ based OTOCs in images (a) and (b), respectively.  The parameter values we use are $N=100$, $W = J^a = 1$ and $\lambda \approx -13.3$, and these numbers give, for the two different spin states of the qubit, the critical energies $E_c = -200$ and $E_c = 0$.  Both Fig.\ \ref{fig:ESotoc}(a) and Fig.\ \ref{fig:ESotoc}(b) show $\overline{F}_n$ is sensitive to the critical energies and in particular, in image (b), using $\hat{\sigma}_z$ qualitatively reproduces the results from Ref. \cite{wang18}.  The black circles and orange triangles represent even and odd parity states, respectively, and are used to show in the symmetry broken phase all energy eigenstates are degenerate, with the degeneracy being broken at the critical energies.

\section{Experimental Considerations}

As briefly mentioned in Sec.\ \ref{sec:model}, one of the experimental platforms for realizing $\hat{H}_B$ is to use a $^{87}\mathrm{Rb}$ BEC where the two spin states are internal states of the atoms \cite{Zibold10}.  Coupling between the two states can be driven via two-photon transitions and the boson-boson interactions can be controlled through the s-wave scattering length by Feshbach resonance \cite{muessel15}.  The role of the qubit could then be played by a neutral impurity atom with two accessible internal states.  Single neutral atom immersion in a $^{87}\mathrm{Rb}$ BEC has been achieved with a $^{133}\mathrm{Cs}$ atom \cite{spethmann12} where the boson-impurity interactions can also be tuned via Feshbach resonance \cite{perrin09}.  If the impurity-impurity interactions are naturally small or can be actively suppressed, then multiple impurities can be used at once thereby giving a larger signal and opening up the possibility of performing multiple measurements without having to reset the system for each measurement.  The suppression of such interactions has been seen experimentally in Bose-Fermi mixtures \cite{roati02,inouye04,gunter06,zaccanti06}.

\begin{figure}[t!]
        \includegraphics[width=0.8\columnwidth]{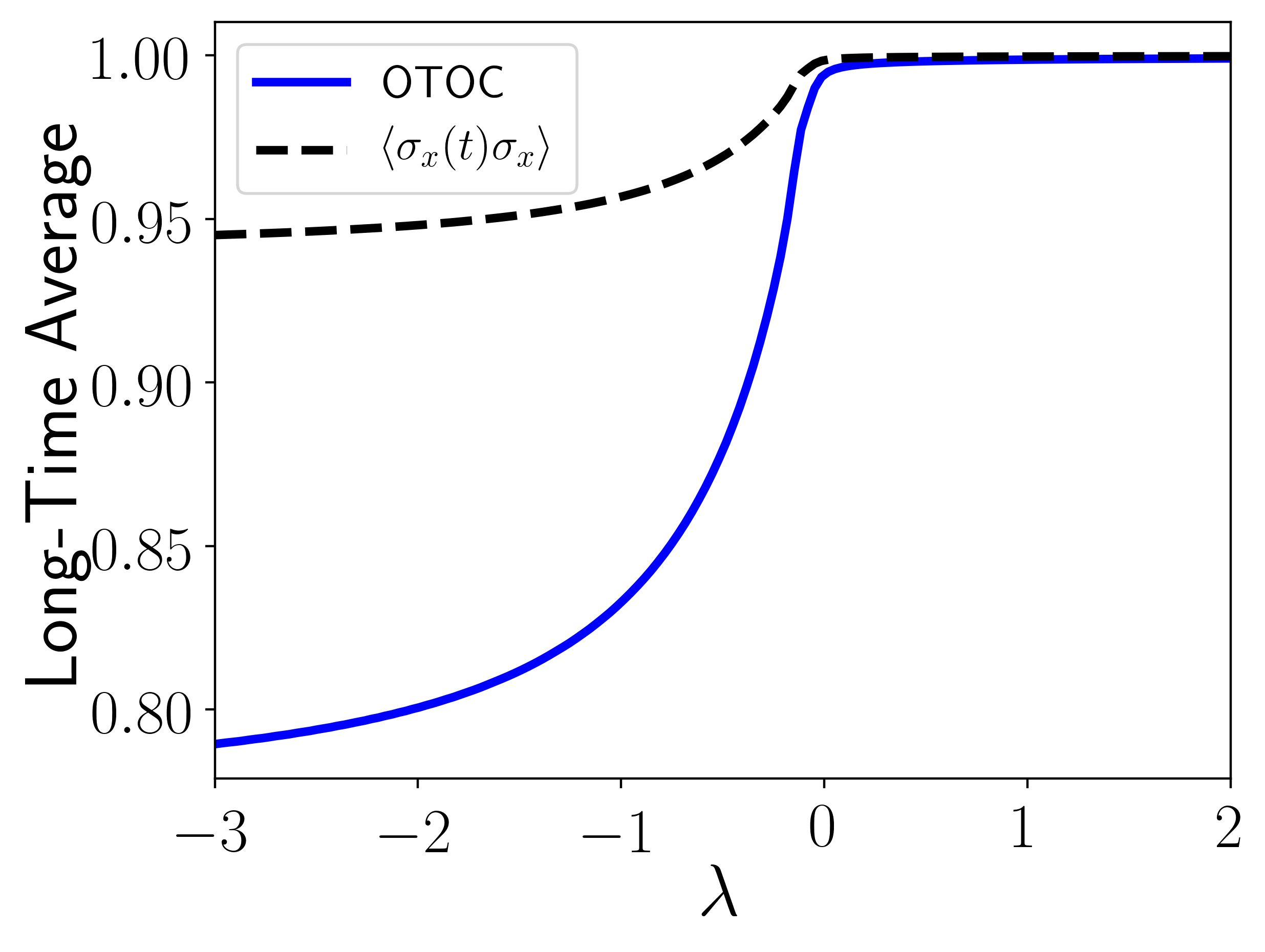} 
        \caption{A comparison of the long-time average of the OTOC $\overline{F}$ (blue, solid) given in Eq.\ (\ref{eq:simpleotocs}) with that of the two-point correlation function  $\overline{\langle \hat{\sigma}_x(t) \hat{\sigma}_x \rangle}$ (black, dashed) as a function of $\lambda$.  In both cases the initial state is chosen to be $\vert \Psi (0) \rangle =  \vert \psi_0 \rangle_B \otimes \left ( \vert + \rangle _Q + \vert - \rangle_Q \right )/\sqrt{2}$ which is a $\hat{\sigma}_{x}$ eigenstate and not only simplifies the calculations but also leads to a relatively simple  experimental protocol. As can be seen, the signature of the QPT is much stronger in the OTOC than the two-point correlation function. The parameter values are $W = J^a = 1$ and $N = 200$.}
\label{fig:twopoint}
\end{figure}

Looking at Eq.\ (\ref{eq:otoc}) it isn't immediately clear how one would go about measuring an OTOC in an experiment.  However, using the OTOC of the qubit rather than the bosons does allow for some flexibility in this regard if the initial state is chosen wisely.  It turns out that the initial product state $\vert \Psi (0) \rangle =  \vert \psi_0 \rangle_B \otimes \left ( \vert + \rangle _Q + \vert - \rangle_Q \right )/\sqrt{2}$, which is easily realized in experiments, simplifies the OTOC function significantly and yet it is still able to properly diagnose the QPT.  In Fig.\ \ref{fig:twopoint} we plot $\overline{F}$ (blue, solid) using $\vert \Psi(0) \rangle$ where we can see it has a sudden drop at the critical point $\lambda = 0$.  The coherent superposition of the two qubit states can be generated by first preparing the qubit in the $\vert - \rangle$ state, then applying a $\pi/2$ pulse. (For large $N$, the ground state of the BEC can be approximated as a spin coherent state of spin-$N/2$, $\vert \psi_0 \rangle_B = \vert \theta_0 , \phi_0 \rangle$, which is regularly prepared in BEC experiments:  the angles define states on the generalized Bloch sphere where $\theta_0$ is the angle with respect to the positive $z$-direction and $\phi_0$ is the azimuthal angle with respect to the positive $x$-direction.  In the normal phase $\theta_0  = \pi/2$, $\phi_0 = 0$ and in the symmetry broken phase $(\theta_0)_\pm = \arccos \left [ \pm \sqrt{ 1- \Lambda^{-2} } \right ]$ and $\phi_0 = 0$ \cite{Zibold10,lee12}.)  Because the initial state $\vert \Psi (0) \rangle$ is an eigenstate of $\hat{\sigma}_x$ the OTOC can be written as 
\begin{equation}
F(t) = \langle \Psi (t) \vert \hat{\sigma}_x \vert \Psi(t) \rangle
\label{eq:simpleotocs}
\end{equation}
where $\vert \Psi(t) \rangle  = e^{i \hat{H}_+ t} e^{-i \hat{H}_- t} \vert \Psi(0) \rangle$ and $\hat{H}_{\pm} = \hat{H}_B + \hat{H}_Q \pm \hat{H}_{I}$.  A derivation of Eq.\ \eqref{eq:simpleotocs} can be found in Appendix \ref{app:Eq21}.  Thus, the  OTOC simplifies to the expectation value of the qubit spin.  However, there remains one major obstacle which plagues the experimental measurement of all OTOCs which is the requirement of backward time propagation, in this case from the operator $\hat{H}_+$.  The difficulty comes from the fact that in order to produce backward time propagation the signs in front of all of the necessary operators need to be flipped, which in our case are the Hamiltonians $\hat{H}_B$ and $\hat{H}_Q$.  Luckily, the boson-boson interactions can be made to be attractive or repulsive via Feshbach resonance and the sign of the coupling parameters between internal states of the bosons and the qubit can also be flipped by putting a $\pi$-phase shift on the two-photon coupling source \cite{muessel15}.  In fact, the OTOC of a $^{87}\mathrm{Rb}$ BEC formed by a lattice in momentum space has been measured experimentally \cite{meier19} where the necessary forward and backward time propagation was achieved through phase shifts in the tunnelling between states.  Finally, the position of the impurity can be measured through fluorescence and the population of a specific spin state can be determined by a state-selective light pulse to remove the populations of all other states \cite{schmidt18}.

\section{Discussion \& Conclusions}

The use of an imbedded qubit to measure the quantum properties of a host many-particle system is appealing for many reasons, not least because the qubit is simple, has a well known spectrum and states which are therefore easy to prepare and address, and it can naturally allow weak measurements that minimally disturb the larger system. In this work, we have shown that despite the simplicity of a single qubit, the time-dependence of a certain four point correlation function---its OTOC---contains the necessary information to characterize both ground state QPTs and excited state QPTs.  Furthermore, the qubit OTOC is able to extract finite size critical scaling exponents from the system if they are coupled for long enough times, however, for short times it fails to extract the dynamical critical exponent because it does not have time to build up strong correlations with the larger system.   We also showed through the PR and the OTO commutator that the effect of the qubit on the larger system is to induce weak information scrambling and chaos.  Our work confirms previous results where weak chaos was found in the level spacing statistics of the eigenenergy values.  In some cases $\overline{F}$ can detect QPTs when two-point correlation functions fail to \cite{heyl18}.  This is not the case for our system, however, we show $\overline{F}$ is much more sensitive than $\overline{\langle \hat{\sigma}_x (t) \hat{\sigma}_x \rangle}$ in detecting the QPT as seen in Fig.\ \ref{fig:twopoint}.

Although in our system the interactions are infinite range resulting in no spatial degrees of freedom, there is also a sense of locality in that the operators we use in Eq.\ (\ref{eq:otoc}) belong to a single particle.  Thus, our system may act as a bridge between OTOCs in infinite range models and the heavily studied OTOCs in lattice systems where the operators used in Eq.\ (\ref{eq:otoc}) are single site operators \cite{Swingle2018,Lin2018,Riddell2019a,Riddell2019b}.  As previously mentioned, OTOCs have been used to calculate the quantum analogue of a system's Lyapunov exponent(s), $\lambda_L$.  With the introduction of the qubit, our system straddles the line between chaos and regular behavior where, like in the Dicke model, in the mean-field limit the dynamics is fully chaotic in the symmetry broken state but regular in the normal state. Thus, an interesting question for future work is whether the Lyapunov exponent can be calculated from the qubit OTOC for our system as well.  In this direction we note that some infinite range models with disorder, such as the Sachdev-Ye-Kitaev model \cite{sachdev93}, have a holographic duality to black holes \cite{maldacena16,jensen16} and share the property of being fast scramblers of information.  In these systems the Lyapunov exponent saturates at the value $2 \pi/ \beta$ \cite{shenker14} where $\beta$ is the inverse temperature, so investigations along these lines for our system may be fruitful.      

\acknowledgments  \textbf{Acknowledgements} We thank the Natural Sciences and Engineering Research Council of Canada (NSERC) for funding.

\appendix

\section{\label{app:Eq21}Derivation of Eq.\ \eqref{eq:simpleotocs}}

To derive Eq.\ \eqref{eq:simpleotocs} we start with the OTOC equation for $\hat{A} = \hat{B} = \hat{\sigma}_x$

\begin{equation}
F(t) = \langle \Psi(0) \vert e^{i\hat{H}_+ t} \hat{\sigma}_x e^{-i\hat{H}_+ t} \hat{\sigma}_x e^{i\hat{H}_+ t} \hat{\sigma}_x e^{-i\hat{H}_+ t} \hat{\sigma}_x \vert \Psi(0) \rangle
\label{eq:eq1}
\end{equation}
where $\vert \Psi (0) \rangle =  \vert \psi_0 \rangle_B \otimes \left ( \vert + \rangle _Q + \vert - \rangle_Q \right )/\sqrt{2}$ which is a product state of the ground states of $\hat{H}_B$ and $\hat{H}_Q$, respectively, and $\hat{H}_\pm = \hat{H}_B + \hat{H}_Q \pm \hat{H}_I$.  First, we use the identity matrix, $\hat{\sigma}_x \hat{\sigma}_x = 1_Q$, and combine it with the fact that $\vert \Psi(0) \rangle$ is an eigenstate of $\hat{\sigma}_x$ with an eigenvalue of $+1$ to make the following transformation of the bra in Eq.\ \eqref{eq:eq1}: $\langle \Psi(0) \vert \to \langle \Psi(0) \vert \hat{\sigma}_x$ giving

\begin{equation}
F(t) = \langle \Psi(0) \vert \hat{\sigma}_x e^{i\hat{H}_+ t} \hat{\sigma}_x e^{-i\hat{H}_+ t} \hat{\sigma}_x e^{i\hat{H}_+ t} \hat{\sigma}_x e^{-i\hat{H}_+ t} \hat{\sigma}_x \vert \Psi(0) \rangle \, .
\label{eq:eq1}
\end{equation}
For a general function of Pauli matrices, $f(\hat{\sigma}_x, \hat{\sigma}_y, \hat{\sigma}_z)$,

\begin{equation}
\hat{\sigma}_x f(\hat{\sigma}_x, \hat{\sigma}_y, \hat{\sigma}_z) \hat{\sigma}_x = f(\hat{\sigma}_x,- \hat{\sigma}_y, -\hat{\sigma}_z) \, ,
\label{eq:eq2}
\end{equation}
so $\hat{\sigma}_x e^{-i\hat{H}_+ t} \hat{\sigma}_x = e^{-i\hat{H}_- t}$ since $\hat{H}_I$ contains a factor of $\hat{\sigma}_z$ ($\hat{H}_Q$ remains unchanged because it has a factor of $\hat{\sigma}_x$).  Applying this transformation to Eq.\ \eqref{eq:eq2} results in Eq.\ \eqref{eq:simpleotocs}

\begin{eqnarray}
F(t) &=&  \langle \Psi(0) \vert e^{i\hat{H}_- t} e^{-i\hat{H}_+ t} \hat{\sigma}_x e^{i\hat{H}_+ t} e^{-i\hat{H}_- t} \vert \Psi(0) \rangle \nonumber \\
&=& \langle \Psi(t) \vert \hat{\sigma}_x \vert \Psi(t) \rangle 
\end{eqnarray}
where $\vert \Psi(t) \rangle  = e^{i \hat{H}_+ t} e^{-i \hat{H}_- t} \vert \Psi(0) \rangle$.


\begin{thebibliography}{8}

\bibitem{bloch08}{I. Bloch, J. Dalibard, and W. Zwerger, Rev. Mod. Phys. \textbf{80}, 885 (2008).}

\bibitem{blatt12}{R. Blatt and C. F. Roos, Nat. Phys. \textbf{8}(4), 227 (2012).}

\bibitem{georgescu14}{I. M. Georgescu, S. Ashhab, and F. Nori, Rev. Mod. Phys. \textbf{86}, 153 (2014).}

\bibitem{prufer18}{M. Pr\"{u}fer, P. Kunkel, H. Strobel, S. Lannig, D. Linnemann, C. M. Schmied, J. Berges, T. Gasenzer, and M. K. Oberthaler, Nature \textbf{563}, 217 (2018).}

\bibitem{zhang17}{J. Zhang, P. W. Hess, A. Kyprianidis, P. Becker, A. Lee, J. Smith, G. Pagano, I. D. Potirniche, A. C. Potter, A. Vishwanath, N. Y. Yao, and C. Monroe, Nature \textbf{543}, 217 (2017).}

\bibitem{choi17}{S. Choi, J. Choi, R. Landig, G. Kucsko, H. Zhou, J. Isoya, F. Jelezko, S. Onoda, H. Sumiya, V. Khemani, C. von Keyserlingk, N. Y. Yao, E. Demler, and M. D. Lukin, Nature \textbf{543}, 221 (2017).}

\bibitem{Hung19}{C.-L. Hung, V. Gurarie, C. Chin, Science \textbf{341}, 1213 (2019).}

\bibitem{steinhauer14}{J. Steinhauer, Nat. Phys. \textbf{10}, 864 (2014); J. Steinhauer, Nat. Phys. \textbf{12}, 959 (2016).}

\bibitem{yao18}{N. Y. Yao, F. Grusdt, B. Swingle, M. D. Lukin, D. M. Stamper-Kurn, J. E. Moore, and E. A. Demler, arXiv:1607.01801 (2016).}

\bibitem{swingle16}{B. Swingle, G. Bentsen, M. Schleier-Smith, and P. Hayden, Phys. Rev. A \textbf{94}, 040302(R) (2016).}

\bibitem{bohrdt17}{A. Bohrdt, C. B. Mendl, M. Endres, and M. Knap, New J. Phys. \textbf{19}, 063001 (2017).}

\bibitem{swan18}{R. J. Lewis-Swan, A. Safavi-Naini, J. J. Bollinger, and A. M. Rey, Nat. Comm. \textbf{10}, 1581 (2019).}

\bibitem{roberts15}{D. A. Roberts and D. Stanford, Phys. Rev. Lett. \textbf{115}, 131603 (2015).}

\bibitem{zhu16}{G. Zhu, M. Hafezi, and T. Grover, Phys. Rev. A \textbf{94}, 062329 (2016).}

\bibitem{kukuljan17}{I. Kukuljan, S. Grozdanov, and T. Prosen, Phys. Rev. B \textbf{96}, 060301 (2017).}

\bibitem{rozenbaum17}{E. B. Rozenbaum, S. Ganeshan, and V. Galitski, Phys. Rev. Lett. \textbf{118}, 086801 (2017).}

\bibitem{cotler18}{J. S. Cotler, D. Ding, G. R. Penington, Ann. Phys. \textbf{396}, 318-333 (2018).}

\bibitem{rozenbaum18}{E. B. Rozenbaum, S. Ganeshan, and V. Galistski, arXiv:1801.10591 (2018).}

\bibitem{kurchan18}{J. Kurchan, J. Stat. Phys. \textbf{171}, 965 (2018).}

\bibitem{chen18}{X. Chen and T. Zhou, arXiv:1804.08655 (2018).}

\bibitem{mata18}{I. G.-Mata, M. Saraceno, R. A. Jalabert, A. J. Roncaglia, and D. A. Wisniacki, Phys. Rev. Lett. \textbf{121}, 210601 (2018).}

\bibitem{carlos18}{ J. C. Carlos, B. L.-del-Carpio, M. A. B.-Magnani, P. Stransky, S. L.-Hernandez, L. F. Santos, and J. G. Hirsch, arXiv: 1807.10292 (2018).}

\bibitem{jalabert18}{R. A. Jalabert, I. G.-Mata, and D. A. Wisniacki, arXiv:1808.04383 (2018).}

\bibitem{hamazaki18}{R. Hamazaki, K. Fujimoto, and M. Ueda, arXiv:1807.02360 (2018).}

\bibitem{herrera18}{E. J. T.-Herrera, A. M. G.-Garcia, and L. F. Santos, Phys. Rev. B \textbf{97}, 060303 (2018).}

\bibitem{he17}{R.-Q. He and Z.-Y. Lu, Phys. Rev. B \textbf{95}, 054201 (2017).}

\bibitem{chen17}{X. Chen, T. Zhou, D. A. Huse, and E. Fradkin, Ann. Phys. (Berlin) \textbf{529}, 1600332 (2017).}

\bibitem{huang17}{Y. Huang, Y.-L. Zhang, and X. Chen, Ann. Phys. (Berlin) \textbf{529}, 1600318 (2017).}

\bibitem{dag18}{C. B. Da\v{g} and L.-M. Duan, arXiv:1807.11085 (2018).}

\bibitem{fan17}{R. Fan, P. Zhang, H. Shen, and H. Zhai, Science Bulletin \textbf{62}, 707 (2017).}

\bibitem{hosur16}{P. Hosur, X.-L. Qi, D. A. Roberts, and B. Yoshida, J. High Energy Phys. 02 (2016) 004.}

\bibitem{heyl18}{M. Heyl, F. Pollman, and B. Dora, Phys. Rev. Lett. \textbf{121}, 016801 (2018).}

\bibitem{sun19}{Z.-H. Sun, J.-Q. Cai, Q.-C. Tang, Y. Hu, and H. Fan, arXiv:1811.11191 (2019).}

\bibitem{shen17}{H. Shen, P. Zhang, R. Fan, and H. Zhai, Phys. Rev. B \textbf{96}, 054503 (2017).}

\bibitem{maldacena16}{J. Maldacena, D. Stanford, and Z. Yang, Prog. Theor. Exp. Phys. \textbf{12}, C104 (2016).}

\bibitem{wei18}{K. X. Wei, C. Ramanathan, and P. Cappellaro, Phys. Rev. Lett. \textbf{120}, 070501 (2018).}

\bibitem{garttner17}{M. G\"{a}rttner, J. G. Bohnet, A. Safavi-Naini, M. L. Wall, J. J. Bollinger, and A. M. Rey, Nat. Phys. \textbf{13}, 781 (2017).}

\bibitem{li17}{J. Li, R. Fan, H. Wang, B. Ye, B. Zeng, H. Zhai, X. Peng, and J. Du, Phys. Rev. X \textbf{7}, 031011 (2017).}

\bibitem{landsman19}{K. A. Landsman, C. Figgatt, T. Schuster, N. M. Linke, B. Yoshida, N. Y. Yao, and C Munroe, Nature \textbf{567}, 61 (2019).}

\bibitem{meier19}{E. J. Meier, J. Ang'ong'a, F. A. An, and B. Gadway, Phys. Rev. A \textbf{100}, 013623 (2019).}

\bibitem{chaudhuri19}{S. Chaudhuri and R. Loganayagam, J. High Energ. Phys.  07 (2019) 006.}

\bibitem{mulansky11}{F. Mulansky, J. Mumford, and D. H. J. O'Dell, Phys. Rev. A \textbf{84}, 063602 (2011).}


\bibitem{Harty14}{T. P. Harty, D. T. C. Allcock, C. J. Ballance, L. Guidoni, H. A. Janacek, N. M. Linke, D. N. Stacey, and D. M. Lucas, High-Fidelity Preparation, Gates, Memory, and Readout of a Trapped-Ion Quantum Bit, Phys. Rev. Lett. 113, 220501 (2014).}

\bibitem{Lanyon17}{B. P. Lanyon, C. Maier, M. Holz\"{a}pfel, T. Baumgratz, C. Hempel, P. Jurcevic, I. Dhand, A. S. Buyskikh, A. J. Daley, M. Cramer, M. B. Plenio, R. Blatt, and C. F. Roos,   Nature Phys. \textbf{13}, 1158 (2017).}

\bibitem{bakr10}{W.S. Bakr, A. Peng, M. E. Tai, R. Ma, J. Simon, J. I. Gillen, S. F\"{o}lling, L. Pollet, and M. Greiner, Science \textbf{329}, 547 (2010).}

\bibitem{sherson10}{J. F. Sherson, C. Weitenberg, M. Endres, M. Cheneau, I. Bloch, and S. Kuhr, Nature \textbf{467}, 68 (2010).}

\bibitem{zipkes10}{C. Zipkes, S. Palzer, C. Sias, and M. K\"{o}hl, Nature \textbf{464}, 388 (2010).}

\bibitem{schmid10}{S. Schmid, A. H\"{a}rter, and J. H. Denschlag, Phys. Rev. Lett. \textbf{105}, 133202 (2010).}

\bibitem{Pinkas19}{M. Pinkas, Z. Meir, T. Sikorsky, R. Ben-Shlomi, N. Akerman and R. Ozeri, arXiv:1907.12815}

\bibitem{spethmann12}{N. Spethmann, F. Kindermann, S. John, C. Weber, D. Meschede, and A. Widera, Phys. Rev. Lett. \textbf{109}, 235301 (2012).}

\bibitem{schmid18}{F. Schmid, D. Mayer, Q. Bouton, D. Adam, T. Lausch, N. Spethmann, and A. Widera, Phys. Rev. Lett. \textbf{121}, 130403 (2018).}

\bibitem{hohmann17}{M. Hohmann, F. Kinderman, T. Lausch, D. Mayer, F. Schmidt, E. Lutz, and A. Widera, Phys. Rev. Lett. \textbf{118}, 263401 (2017).}

\bibitem{usui18}{A. Usui, B. Bu\v{c}a, and J. Mur-Petit, New J. Phys. \textbf{20}, 103006 (2018).}

\bibitem{elliott16}{T. J. Elliott and T. H. Johnson, Phys. Rev. A \textbf{93}, 043612 (2016).}







\bibitem{kac63}{M. Kac, G. E. Uhlenbeck, and P. C. Hemmer, J. Math. Phys. \textbf{4}, 216 (1963).}

\bibitem{Islam11}{R. Islam, E.E. Edwards, K. Kim, S. Korenblit, C. Noh, H. Carmichael, G.-D. Lin, L.-M. Duan, C.-C. Joseph Wang, J.K. Freericks, and C. Monroe, Nature Comm. \textbf{2}, 377 (2011).}

\bibitem{Britton12}{J. W. Britton, B. C. Sawyer, A. C. Keith, C.-C. J. Wang, J. K. Freericks, H. Uys, M. J. Biercuk, and J. J. Bollinger, Nature (London) \textbf{484}, 489 (2012).}

\bibitem{Islam13}{R. Islam, C. Senko, W.C. Campbell, S. Korenblit, J. Smith, A. Lee, E. E. Edwards, C.-C. J. Wang, J. K. Freericks, and C. Monroe, Emergence and Frustration of Magnetic Order with Variable-Range Interactions in a Trapped Ion Quantum Simulator, Science 340, 583 (2013).}

\bibitem{Jurcevic14}{P. Jurcevic, B. P. Lanyon, P. Hauke, C. Hempel, P. Zoller, R. Blatt, and C.F. Roos, Nature (London) \textbf{511}, 202 (2014).}

\bibitem{Richerme14}{P. Richerme, Z.-X. Gong, A. Lee, C. Senko, J. Smith, M. Foss-Feig, S. Michalakis, A. V. Gorshkov, and C. Monroe, Nature (London) \textbf{511}, 198 (2014).}

\bibitem{Bohnet16}{J. G. Bohnet, B. C. Sawyer, J. W. Britton, M. L. Wall, A. M. Rey, M. Foss-Feig, and J.J. Bollinger, Science \textbf{352}, 1297 (2016).}

\bibitem{Das06}{A. Das, K. Sengupta, D. Sen, and B.K. Chakrabarti, Phys. Rev. B 74, 144423 (2006).}

\bibitem{albeiz05}{M. Albiez, R. Gati, J. F\"{o}lling, S. Hunsmann, M. Cristiani, and M. K. Oberthaler, Phys. Rev. Lett. \textbf{95}, 010402 (2005).}

\bibitem{levy2007}{S. Levy, E. Lahoud, I. Shomroni, and J. Steinhauer, Nature (London) \textbf{449}, 579 (2007).}

\bibitem{Trenkwalder16}{A. Trenkwalder, G. Spagnolli, G. Semeghini, S. Coop, M. Landini, P. Castilho, L. Pezz\`{e}, G. Modugno, M. Inguscio, A. Smerzi, and M. Fattori, Nat. Phys. \textbf{12}, 826 (2016).}

\bibitem{Zibold10}{T. Zibold, E. Nicklas, C. Gross, and M.K. Oberthaler, Phys. Rev. Lett. \textbf{105}, 204101 (2010).}

\bibitem{Rinck2011}{M. Rinck and C. Bruder, Phys. Rev. A \textbf{83}, 023608 (2011).}

\bibitem{mumford14a}{J. Mumford, J. Larson, and D. H. J. O'Dell, Phys. Rev. A \textbf{89}, 023620 (2014).}

\bibitem{mumford14b}{J. Mumford and D. H. J. O'Dell, Phys. Rev. A \textbf{90}, 063617 (2014).}

\bibitem{Dicke54}{R. H. Dicke, Phys. Rev. \textbf{93}, 99 (1954).}

\bibitem{Hepp73}{K. Hepp and E. H. Lieb, Ann. Phys. (NY) \textbf{76}, 360 (1973).}

\bibitem{Wang73}{Y. K. Wang and F. T. Hioe, Phys. Rev. A \textbf{7}, 831 (1973).}

\bibitem{Garraway11}{B. M. Garraway, Phil. Trans. R. Soc. A \textbf{369}, 1137 (2011).}

\bibitem{Dimer07}{F. Dimer, B. Estienne, A. S. Parkins, and H. J. Carmichael, Phys. Rev. A 75, 013804 (2007).}

\bibitem{Nagy10}{D. Nagy, G. Konya, G. Szirmai, and P. Domokos, Phys. Rev. Lett. \textbf{104}, 130401 (2010); D. Nagy, G. Szirmai, and P. Domokos, Phys. Rev. A \textbf{84}, 043637 (2011).}

\bibitem{Bhaseen12}{M. J. Bhaseen, J. Mayoh, B. D. Simons, and J. Keeling, Phys. Rev. A \textbf{85}, 013817 (2012).}

\bibitem{Baumann10}{K. Baumann, C. Guerlin, F. Brennecke, and T. Esslinger, Nature (London) \textbf{464}, 1301 (2010).}

\bibitem{Safavi18}{A. Safavi-Naini, R. J. Lewis-Swan, J. G. Bohnet, M. G\"{a}rttner, K. A. Gilmore, J. E. Jordan, J. Cohn, J. K. Freericks, A. M. Rey, and J. J. Bollinger, Phys. Rev. Lett. \textbf{121}, 040503 (2018).}

\bibitem{lambert04}{N. Lambert, C. Emary, and T. Brandes, Phys. Rev. Lett. \textbf{92} 073602 (2004); C. Emary and T. Brandes, Phys. Rev. Lett. \textbf{90} 044101 (2003); C. Emary and T. Brandes, Phys. Rev. E \textbf{67}, 066203 (2003).}

\bibitem{buonsante12}{P. Buonsante, R. Burioni, E. Vescovi, and A. Vezzani, Phys. Rev. A \textbf{85}, 043625 (2012).}

\bibitem{Sakurai93}{J. J. Sakurai, Modern Quantum Mechanics (Addison-Wesley,
Boston, 1993).}

\bibitem{Gati07}{R. Gati and M. K. Oberthaler, J. Phys. B: At., Mol. Opt. Phys.
\textbf{40}, R61 (2007).}

\bibitem{lipkin65}{H. Lipkin, N. Meshkov, and A. Glick, Nuclear Physics \textbf{62}, 188 (1965).}

\bibitem{dusuel04}{S. Dusuel and J. Vidal, Phys. Rev. Lett. \textbf{92}, 237204 (2004); S. Dusuel and J. Vidal, Phys. Rev. B \textbf{71}, 224420 (2005).}





\bibitem{buijsman17}{W. Buijsman, V. Gritsev, and R. Sprik, Phys. Rev. Lett. \textbf{118}, 080601 (2019).}



\bibitem{dag19}{C. B. Da\v{g}, K. Sun, and L.-M. Duan, arXiv:1902.05041 (2019).}

\bibitem{borgonovi19}{F. Borgonovi, F. M. Izrailev, and L. F. Santos, arXiv:1903.09175 (2019).}

\bibitem{wei19}{B.-B. Wei, G. Sun, and M.-J. Hwang, arXiv:1906.00533 (2019).}

\bibitem{wang18}{Q. Wang and F. P\'{e}rez-Bernal, arXiv:1812.01920 (2018).}

\bibitem{muessel15}{W. Muessel, H. Strobel, D. Linnemann, T. Zibold, B. Juli\'{a}-D'{i}az, and M. K. Oberthaler, Phys. Rev. A \textbf{92}, 023603 (2015).}

\bibitem{perrin09}{H. Perrin, EPJ-Special Topics \textbf{172}, 37 (2009).}

\bibitem{roati02}{G. Roati, F. Riboli, G. Modugno, and M. Inguscio, Phys. Rev. Lett. \textbf{89}, 150403 (2002).}

\bibitem{inouye04}{S. Inouye, J. Goldwin, M. L. Olsen, C. Ticknor, J. L. Bohn, and D. S. Jin, Phys. Rev. Lett. \textbf{93}, 183201 (2004).}

\bibitem{gunter06}{K. G\"{u}nter, T. St\"{o}ferle, H. Moritz, M. K\"{o}hl, and T. Esslinger, Phys. Rev. Lett. \textbf{96}, 180402 (2006).}

\bibitem{zaccanti06}{M. Zaccanti, C. D'Errico, F. Ferlaino, G. Roati, M. Inguscio, and G. Modugno, Phys. Rev. A \textbf{74}, 041605 (2006).}

\bibitem{lee12}{C. Lee, J. Huang, H. Deng, H. Dai, and J. Xu, Front. Phys. \textbf{7}, 109 (2012).}

\bibitem{schmidt18}{F. Schmidt, D. Mayer, Q. Bouton, D. Adam, T. Lausch, N. Spethmann, and A. Widera, Phys. Rev. Lett. \textbf{121}, 130403 (2018).}

\bibitem{Swingle2018}{S. Xu and B. Swingle, arXiv:1802.00801.}
	
	\bibitem{Lin2018}{C.-J. Lin and O. I. Motrunich, arXiv:1801.01636.}
	 
	 \bibitem{Riddell2019a}{J. Riddell and E. S. S\o{}rensen, Out-of-time ordered correlators and entanglement growth in the random-field XX spin chain, Phys. Rev. B \textbf{99}, 054205 (2019).}
	 
	 \bibitem{Riddell2019b}{J. Riddell and E. S. S\o{}rensen, arXiv:1908.03292}


\bibitem{sachdev93}{S. Sachdev and J. Ye, Phys. Rev. Lett. \textbf{70}, 3339 (1993).}

\bibitem{jensen16}{K. Jensen, Phys. Rev. Lett. \textbf{117}, 111601 (2016).} 

\bibitem{shenker14}{S. H. Shenker and D. Stanford, J. High Energy Phys. \textbf{03}, 067 (2014).}














\end{thebibliography}
\end{document}